\documentclass[11.5pt]{article}
\usepackage{}
\usepackage{wrapfig}
\usepackage{cite,amssymb,amsfonts,hangcaption,amsmath,here,color,bm} 
\usepackage{graphicx,here}  

\renewcommand{\theequation}{\arabic{section}.\arabic{equation}}

\makeatletter
\@addtoreset{equation}{section}
\@addtoreset{footnote}{section} 
\makeatother

\makeatletter
\renewcommand\appendix{
\par%\newpage
\setcounter{section}{0}%
\setcounter{subsection}{0}%
\gdef\thesection{Appendix \@Alph\c@section }
\renewcommand{\theequation}
{\Alph{section}.\arabic{equation}}
} 
\makeatother

\makeatletter
\def\eqnarray{ \stepcounter{equation} \let\@currentlabel=\theequation
\global\@eqnswtrue
\global\@eqcnt\z@
\tabskip\@centering
\let\\=\@eqncr
$$\halign to \displaywidth\bgroup\@eqnsel\hskip\@centering
$\displaystyle\tabskip\z@{##}$&\global\@eqcnt\@ne
\hfil$\displaystyle{{}##{}}$\hfil
&\global\@eqcnt\tw@$\displaystyle\tabskip\z@{##}$\hfil
\tabskip\@centering&\llap{##}\tabskip\z@\cr}
\makeatother

\makeatletter
\def\@arrayacol{\edef\@preamble{\@preamble \hskip .5\arraycolsep}}
\def\array{\let\@acol\@arrayacol \let\@classz\@arrayclassz
\let\@classiv\@arrayclassiv \let\\\@arraycr\def\@halignto{}\@tabarray}
\makeatother
\makeatletter
\newcounter{subeqncnt}
\def\thesubeqncnt{\alph{subeqncnt}}
\def\subequations{\begingroup%
\stepcounter{equation}\edef\@tempa{\theequation}%
\let\c@equation\c@subeqncnt\c@subeqncnt\z@
\edef\theequation{\@tempa\noexpand\thesubeqncnt}}

\makeatother
\newcommand{\captionfonts}{\small}
\makeatletter % Allow the use of @ in command names
\long\def\@makecaption#1#2{%
\vskip\abovecaptionskip
\sbox\@tempboxa{{\captionfonts #1: #2}}%
\ifdim \wd\@tempboxa >\hsize
{\captionfonts #1: #2\par}
\else
\hbox to\hsize{\hfil\box\@tempboxa\hfil}%
\fi
\vskip\belowcaptionskip}
\makeatother % Cancel the effect of \makeatletter

\newcommand{\Tr}{\mathop{\rm Tr}\nolimits}

\newcommand{\Exp}{\mbox{Exp}}

\newcommand{\be}{\begin{equation}}
\newcommand{\ee}{\end{equation}}
\newcommand{\bea}{\begin{eqnarray}}
\newcommand{\eea}{\end{eqnarray}}

\newcommand{\nq}{\mathfrak{q}}
\newcommand{\nw}{\mathfrak{w}}
\newcommand{\nmu}{\mathfrak{u}}

\begin{document}

\setlength{\baselineskip}{7mm}
\begin{titlepage}
\begin{flushright}
{\tt CAS-KITPC/ITP-349} \\
%{\tt arXiv:****.****} \\
%{\tt December 2012} \\
\end{flushright}

%\vspace{1cm}

%\begin{center}

\setlength{\baselineskip}{9mm}

\begin{center}

\vspace*{15mm}
{\LARGE
Hydrodynamics and transport coefficients 
in an infrared-deformed soft-wall AdS/QCD
model at finite temperature
}

\vspace{8mm}
{\large
%{\sc{Ling-Xiao Cui}}
%%$^\dagger$
%
{\sc{Shingo Takeuchi}}
%$^\dagger$
 and
{\sc{Yue-Liang Wu}}
%$^\dagger$
}

\setlength{\baselineskip}{0mm}

\vspace*{12mm}
{\large
%$\dagger$
{\it{State Key Laboratory of Theoretical Physics}}\\
{\it{Kavli Institute for Theoretical Physics China (KITPC)}}\\
{\it{Institute of Theoretical Physics, Chinese Academy of Sciences, Beijing 100190, China}} \\
}

\end{center}

\vspace{1cm}

{\large
\begin{abstract}
{
We extend an 
infrared-deformed soft-wall 
anti de-Sitter/QCD model at zero temperature to a model at finite temperature and perform hydrodynamics. 
To have the infalling
boundary condition to make the hydrodynamic analysis possible, we treat the infrared energy scale
factor in our metric as a temperature-depending parameter. Then, 
by carrying out the hydrodynamic
analysis, we compute the transport coefficients, the diffusion constant, 
and the shear viscosity 
through the linear response theory.
}
\end{abstract}
}

\end{titlepage}

%============================================
\section{Introduction}
\label{Chap:Intro}
%============================================

To carry out the analysis 
in the strongly coupled region would be the key of many important unsolved
problems in contemporary high-energy theoretical physics. 
One of these problems is the low-energy dynamics of QCD. 
Actually, many ideas for the nonperturbative analysis of QCD have been proposed. Among
these, there are two interesting classes, the lattice gauge theory~\cite{Lattice} 
 and holography~\cite{AdSCFT,GKPW} (gauge/gravity correspondence, anti-de-Sitter (AdS)/QCD, etc.
%To carry out the analysis 
%in the strongly coupled region 
%would be the key of many important unsolved problems 
%in the contemporary high energy theoretical physics. 
%%---
%One of these problems is the low energy dynamics of quantum chromodynamics~(QCD).  
%Actually, 
%many ideas  
%for the non-perturbative analysis of QCD have been proposed. 
%%---
%Among these, 
%there are two interesting classes, 
%the lattice gauge theory~\cite{Lattice} and holography~\cite{AdSCFT,GKPW}
%~(gauge/gravity correspondence, AdS/QCD, etc).
\newline

In the lattice gauge theory, 
Minkowski space-time in the original theory is replaced with a finite volume
Euclidian discretized space-time with an analytic continuation for the time direction to the imaginary
time direction. As a result, the degree of freedom of theories becomes finite, and the nonperturbative 
numerical analysis for the action itself, which is a Monte Carlo simulation, becomes
available. However, Monte Carlo simulation is plagued by a notorious problem named as the sign problem, 
when fermions are involved in the Monte Carlo. Currently, no fundamental means 
to overcome the sign problem has been invented yet, and the current analyses 
in the lattice are always carried out by getting around it~\cite{Ejiri:2010vm}.

%In the lattice gauge theory, 
%Minkowski space-time  
%in the original theory 
%is replaced
%with a finite volume Euclidian discretized space-time 
%with analytic continuation for time direction
%to imaginary time direction.
%%---
%As a result, the degree of freedom of theories becomes finite,
%and the non-perturbative numerical analysis for the action itself,
%which is Monte Carlo simulation, becomes available.
%%---
%However, Monte Carlo simulation is plagued 
%by a notorious problem 
%named as the sign problem,
%when fermions are involved in the Monte Carlo.
%%---
%Currently, 
%no fundamental mean 
%to overcome the sign problem  
%has been invented yet,
%and the current analysis  
%in the lattice are always carried out
%by getting around it\cite{Ejiri:2010vm}.

On the other hand, it is well known that the holography is a duality
between the strongly coupled field theories and the weakly coupled gravities.
%---
It originates in the superstring theory, 
in which the quantized open and closed strings
at low energy can be identified with particles
known in field theories and gravities~\cite{StringParticle}.
%---
As a result, behaviors of the low-energy open strings
describe the supersymmetric gauge theory.
%---
Then, as an important matter in the correspondence,
the $U(N)$ supersymmetric gauge theory arises
on $N$ overlapped D-branes from the open strings
sticking to these $N$ overlapped D-branes
at low energy.
%---
On the other hand, D-branes can be identified with black branes
in the supergravity~\cite{Dbraneblackbrane}.

Thus, one has two ways to describe a low-energy D-branes.
%---
As a result, it is known that one can conjecture the duality
between $p$-dimensional large-$N$ supersymmetric $SU(N)$ gauge theory
with large 't Hooft coupling and near-horizon geometries of the black $p+1$-brane
in the condition that 
the quantum effect of gravity and the length of the string can be neglected.
%---
Although there is no exact proof for this correspondence
at this moment, it is particularly expected that the duality between
the ${\cal N}=4$ four-dimensional large-$N$ $SU(N)$ supersymmetric gauge theory~\cite{AdSCFT,GKPW}
and the five-dimensional anti de-Sitter space is valid.

One of the great advantages in the gauge/gravity correspondence 
compared with the lattice gauge theory would be that
it is irrelevant to the problem arising when one involves fermions
like the sign problem, 
%---
because 
the main analyses are carried out 
in analytic ways 
in the weakly coupled gravity side.

However, 
the current gauge/gravity correspondence also has a problem.
It is that the dual gauge theories are always nonrealistic
as long as the gravity side is a solution~(top-down model).
%---
Because of this, 
many results in the field theory side
in the current gauge/gravity correspondence
are the ones independent of detail of theories
or no more than qualitative ones just in supersymmetric models. 
%---
On the other hand, 
once getting away from the study
based on a solution  in the gravity side, constructing holographic models
in bottom-up way is also conducted energetically~\cite{Erlich:2005qh,Karch:2006pv,AdS-QCD_Review}.

Anyway, 
the point that the gauge/gravity correspondence can be irrelevant
to the notorious problem in the treatment of fermions
in the lattice gauge theory 
would be one of the great advantages.
%---
For this reason, 
the low-energy dynamics of QCD has recently been studied intensively
in the framework of the gauge/gravity correspondence, 
and this is the motivation of the soft-wall AdS/QCD model, 
which is a kind of the holographic bottom-up model~\cite{Karch:2006pv,AdS-QCD_Review}.
\newline

Recently, 
we have done the extension of an IR-deformed AdS/QCD model~\cite{Sui:2009xe} 
to the finite temperature system~\cite{Cui:2011wb,Cui:2011ag}, 
in which the deformed bulk vacuum and potential term have been introduced
for the scalar field to satisfy the equations of motion.  
%---
This is because 
if one straightforwardly extends the model 
to the finite temperature system, 
it turns out that 
the solution of the equation of motion diverges due to the dilaton. 
%---
Only by deforming the bulk vacuum and potential term, 
one can obtain the smooth solution 
for the dilation.
%---
With such a treatment, 
we have examined
the critical temperature of chiral symmetry breaking~\cite{Cui:2011wb}
through the analysis of the quark number susceptibility 
and the meson spectrum~\cite{Cui:2011ag}.
%---
In the analysis of the meson spectrum,  
we have carried out
the numerical analysis of the equation of motion
for the fluctuations on the bulk gravity.
%---
Such a numerical analysis for the mass spectrum can be considered
as a \textcolor{black}{basic} method of analyses in holographic QCD
as well as the hydrodynamics.

One of the crucial points of the hydrodynamics is that
it can be considered to be independent of the detail of theories,
and the transport coefficients are also so independent of the detail of theories,
for which the transport coefficients are the ones
defined in the framework of \textcolor{black}{the hydrodynamic analysis.}
%---
In particular, the ratio 
between the shear viscosity ($\eta$) and the entropy density ($s$) $\eta/s$ 
is the quantity characterizing the actual QCD.
For this reason, the ratio $\eta/s$ has been examined very much in the gauge/gravity correspondence~\cite{Cremonini:2011iq}.
Besides, the holographic hydrodynamics has played
an important role in the long-standing problem
in the causal hydrodynamics~\cite{Natsuume:2007ty}.
and quark-gluon plasma described by the Bjorken flow~\cite{ReviewBjorkenFlow}.
%---

For such circumstances,  
turning to the hydrodynamics,
in this paper,  
we are going to work out the holographic hydrodynamics 
in an IR-deformed AdS/QCD model 
at finite temperature studied~\cite{Cui:2011wb,Cui:2011ag}. 
to perform the interesting studies 
mentioned above in the future. 
\newline

Now,  
we would like to mention 
the organization of this paper. 
%---
In Sec.\ref{Chap:Hydro}, 
we review the hydrodynamics and the transport coefficients
obtained from the linear response theory. 
%---
In Sec.\ref{Chap:AdSQCD model},
we will introduce our holographic model 
and show how the model is extended
to the finite temperature system.
It is shown that,  
if one takes the same way 
as in Ref.\cite{Cui:2011ag},
it turns out that 
the analysis becomes too complicated
to be carried out. 
Therefore, 
in this paper,
we will propose another way,
which is simpler than the one
in Ref.\cite{Cui:2011ag} 
%as the quark-gluon plasma is 
%considered to be above the critical temperature.
%-----
Then, 
it will be seen that 
there are four options
in the numerical calculations for some factors,
while only one of them is physically acceptable. 
%-----
In Sec.\ref{Chap:Preliminaries},
we will sort out the notation
used in this paper.
%before starting the analysis.
%----
In Sec.\ref{Chap:U(1)_baryon_current},
we will carry out the hydrodynamic analysis
for the fluctuation of the balk gravity
dual to the $U(1)$ baryon current 
in the scalar mode.
In the analysis, 
a necessity
to cancel the divergence
at the horizon arises 
as usual.
Our model involves the dilaton which makes the analysis complicated.
However, we will show how to contain it
by exploiting the integral constant.
Then, using the Gubser, Klebanov, Polyakov, and Witten relation~(using GKP-W) relation~\cite{GKPW},
we will read out the retarded Green function
for the $U(1)$ baryon current 
in the scalar mode
and its diffusion constant.
%----
In Sec.\ref{Chap:SU(2)_flavor_current},
we will carry out 
the hydrodynamic analysis
for the fluctuation of the balk gravity
dual to the $SU(2)$ flavor current 
in the vector mode
as well as in Sec.\ref{Chap:U(1)_baryon_current}.
Then, 
using the GKP-W relation~\cite{GKPW},
we will read out the retarded Green function. 
%----
In Sec.\ref{Chap:tensor},
we will carry out the hydrodynamic analysis
for the fluctuation of the tensor mode 
in the balk gravity
dual to the energy-momentum tensor.
Then, using the GKP-W relation~\cite{GKPW},
we can read out the viscosity
with the retarded Green function
and evaluate the ratio between  the viscosity~($\eta$) and entropy density~($s$) as $\eta/s$.
Our summary and conclusions will be presented in Sec.\ref{Chap:Sum}.

%============================================
\section{Brief review on hydrodynamics and linear response theory}
\label{Chap:Hydro}
%============================================

To begin with, 
we would like to describe 
the basic matters
in hydrodynamics and the linear response theory used in this paper.
%---
The description in this section is basically following
the review papers~\cite{review_hydro_linear}.

%---
The hydrodynamics is an effective theory
to describe the macroscopic dynamics
at large distances and time scales.
%---
Conserved quantities are considered
to survive in such large distances and time scales,
and the energy-momentum tensor $T^{\mu\nu}$ is
one of conserved quantities.
%---
The hydrodynamics is formulated
by the hydrodynamic equation 
for conserved quantities
instead of the action principle.
%---
The hydrodynamic equation
for the energy-momentum tensor
is given as
\begin{eqnarray}\label{hydro01}
\partial_\mu T^{\mu\nu} &=& 0,
\end{eqnarray}
where
\begin{eqnarray}\label{Tmn}
T^{\mu\nu} &=& (\varepsilon + P)u^\mu u^\nu + Pg^{\mu\nu} + \tau^{\mu\nu}
\end{eqnarray}
with
energy density $\varepsilon$,
pressure $P$, local fluid velocity $u^\mu$ and
the $\tau^{\mu\nu} $ given as
\begin{eqnarray}
\tau^{\mu\nu} &=& -\eta
\left(
\partial^\mu u^\nu + \partial^\nu u^\mu
-
\frac{2}{3} \eta^{\mu\nu} \partial_\alpha u^\alpha
\right)
-
\zeta \eta^{\mu\nu} \partial_\alpha u^\alpha.
\end{eqnarray}
Here $\eta$ and $\zeta$ mean shear and bulk viscosities, respectively.
%This is the constitutive equation.
In a curved space, it is given as
\begin{eqnarray}\label{consteq}
\tau^{\mu\nu} &=&
-P^{\mu \alpha}P^{\nu \beta}
\left\{
\eta
\left(
\nabla_\alpha u_\beta
+
\nabla_\beta u_\alpha
-
\frac{2}{3}
g_{\alpha \beta}
\nabla   u
\right)
-
\zeta g_{\alpha \beta}
\nabla  u
\right\},
\end{eqnarray}
where $P^{\mu \nu} = g^{\mu \nu} +u^\mu u^\nu$.
In the above, $\eta$ and $\zeta$ are regarded as the transport coefficients.
In what follows we consider the fluid
in the rest frame, $u^\mu=(1,0,0,0)$.

In Eq.(\ref{Tmn}),
substituting $u^\mu=(1,0,0,0)$
and expanding as $g_{\mu \nu}=\eta_{\mu \nu}+h_{\mu \nu}$
~[$\eta_{\mu \nu}={\rm diag}(-1,1,1,1)$ and $h_{\mu \nu}$ means fluctuations],
it turns out that $\delta \langle \tau^{xy} \rangle$ is given as
\begin{eqnarray}\label{dtauxy}
\delta \langle \tau^{xy} \rangle = i \omega \eta h_{xy},
\end{eqnarray}
where we have represented the formula
in the momentum space.
%-----
It is known
from the linear response theory
that
the response of an operator $\mathcal{O}$
for the external field $\phi_{(0)}(k)$
is given as
\begin{equation}
\label{hydro01}
\delta \langle \mathcal{O}(k) \rangle
=
-G^{ \mathcal{O} \mathcal{O}}_{\rm R} \mathcal{O}(k),
\end{equation}
where
$ \displaystyle
G^{ \mathcal{O} \mathcal{O}}_{\rm R}(k)
\equiv
-i \int^\infty_{-\infty} d^4 x e^{-i kx} \langle [\mathcal{O}(t,{\bm x}),\mathcal{O}(0,0)] \rangle \theta(t)
$
is the retarded Green function.
%---
Then, by comparing Eq.(\ref{dtauxy}) with Eq.(\ref{hydro01}),
we can obtain the Kubo formula
with regard to the shear viscosity as
\begin{eqnarray}\label{Kuboeta}
\eta = - \lim_{\omega \to 0}\frac{1}{\omega}\,{\rm Im}\,G_{\rm R}^{xy \ xy},
\end{eqnarray}
where
$ \displaystyle
G^{xy \ xy}_{\rm R}(k)
\equiv
-i \int^\infty_{-\infty} d^4 x e^{-i kx} \langle [T^{xy}(t,{\bm x}),T^{xy}(0,0)] \rangle \theta(t)
$.
%\newline

Next we turn to the diffusion
in the conserved current
and the energy-momentum tensor.
%---
Toward a conserved current $j^\mu$
with $0=\partial_\mu j^\mu$,
the constitutive equation
for the conserved current
is given as
\begin{eqnarray}
{\bm j}
= \rho {\bm u}  - D \nabla j^t
= - D \nabla j^t,
\end{eqnarray}
where $j^t=\rho$ and $D$ mean
the charge density and the diffusion constant respectively,
and we have taken account of the fluid rest frame.
%---
Then, Fick's law
is valid as
\begin{eqnarray}\label{difconst}
\partial_t \rho - D \nabla^2 \rho =0.
\end{eqnarray}
This gives the following dissipation relation
for the charge density:
\begin{eqnarray}
\omega = -i D k^2,
\end{eqnarray}
which will be the pole
in the retarded Green function
in the charge density,
$\displaystyle
G^{tt}_{\rm R}(k)
\equiv
-i \int^\infty_{-\infty} d^4 x e^{-i kx} \\ \langle [\rho(t,{\bm x}),\rho(0,0)] \rangle \theta(t)$.
%---
Next,
we consider the diffusion
in the the energy-momentum tensor $T^{ti}$
with $i=x,y$.
%---
First,
from the constitutive equation
in the curved space,
one can obtain
\begin{eqnarray}
T^{zi}
=
-\frac{\eta}{\varepsilon+P} \partial_z T^{ti}
=
-\eta \partial_z u^i.
\end{eqnarray}
Here,
we have arranged the $k$
along with the $x^3$ axis as $k=(0,0,k)$
for simplicity.
%---
Then,
from $\partial_\mu T^{\mu i} = 0$,
one can obtain
\begin{eqnarray}
\partial_t T^{ti}-\frac{\eta}{\varepsilon+P} \partial_z^2 T^{ti}=0.
\end{eqnarray}
This gives the dissipation relation in the energy-momentum tensor $T^{ti}$,
\begin{eqnarray}
\omega = -i \frac{\eta}{\varepsilon+P} k^2,
\end{eqnarray}
which will be the pole
in the retarded Green function $G_{\rm R}^{ti \ ti}$.

Finally, we write down
the decomposition of
the energy-momentum current
and the $U(1)$ current
under the little group $SO(2)$
toward the $i$ direction ($i=x,y$):
\begin{eqnarray}\label{clsTJ}
&& \mbox{scalar mode: } T_{00},\, T_{03},\, T_{33},\, T^i{}_i \quad {\rm and} \quad J_0,\, J_3,  \nonumber \\
&& \mbox{vector mode: } T_{0i},\, T_{3i}\quad {\rm and} \quad J_i, \nonumber \\
&& \mbox{tensor mode: } T_{ij}-\delta_{ij} T^k{}_k/2.\nonumber
\end{eqnarray}
%Here, the notation in the above follows the one given in Table.(\ref{v1}).

%============================================
\section{IR-deformed AdS/QCD model at finite temperature }
\label{Chap:AdSQCD model}
%============================================

We will start with the following geometry,
which is deformed from Schwarzschild $AdS_5$ black hole geometry
in the IR-region by a factor $\mu_g$ as
\begin{equation}\label{metric_dads_01}
ds^2
=
a^2(z)
\left(
- f(z) dt^2
+ \sum_{i=1}^3 dx^2_i
+ \frac{dz^2}{f(z)}
\right)
\end{equation}
with $a^2(z) = (1/z^2 + \mu_g^2)/l^2$ and $f(z)=1-(z/z_0)^4$
($z_0$ means the location of the horizon)
and we have put the AdS radius $l$ as $1$
in what follows.
The coordinate $z$ is in the relation with the usual coordinate $r$ as $z=1/r$.
This geometry has Hawking temperature $T=1/(\pi z_0)$ 
and is asymptotically $AdS_5$ space-time.
\newline

We will consider
the following $U(1) \times SU_L(2) \times SU_R(2)$ soft-wall model
with the scalar field
on the background (\ref{metric_dads_01}) as
\begin{eqnarray} \label{action01}
S
&=&
\int d^5x \,
\sqrt{-g} e^{-\Phi(z)}
\left(R+12\right)
+
\int d^5x \,
\sqrt{-g} e^{-\Phi(z)}
\Bigg[
-\frac{1}{4g_{\rm U(1)}^2}
F_{MN}F^{MN}
\nonumber \\
&&
+
{\rm{Tr}} \,
\bigg\{
-\frac{1}{4g_{\rm SU(2)}^2}
\Big(
F_{L,MN}F_L^{MN} + F_{R,MN}F_R^{MN}
\Big)
%\right.
%\left.
+ |D_M X(z)|^2
- m_X^2 |X(z)|^2 - \frac{\lambda}{4} |X(z)|^4
\bigg\}
\Bigg],
\end{eqnarray}
where $m_X^2=-3$~\cite{Karch:2006pv,Erlich:2005qh}
and $l=1$.
The trace is performed
for the $SU(2)$ algebra 
mentioned in what follows.
%---
We write $F_{MN}$, $F_{L,MN}$ and $F_{R,MN}$
as
$
F_{MN} 
\equiv 
\partial_M A_N 
- 
\partial_N A_M
$, 
and
$F_{L,MN} 
\equiv 
\partial_M B_{L,N} 
- 
\partial_N B_{L,M} 
-
i 
[B_{L,M},B_{L,N}]
$,
where
$B_{L,M} 
= 
B_{L,M}^a t^a$
with the $SU(2)$ Lie algebra $t^a$ $(a = 1, 2, 3)$,
and now we have skipped describing the $R$ part.
%---
As for the dual operators
for these in the boundary theory,
for example see
the table in Ref.\cite{Erlich:2005qh}.
%---
Using these,
the covariant derivative can be written as
$D_M X = \partial_M X + i (B_{L,M}X - X B_{R,M})$.

We write the bulk vacuum of the scalar field as
%The bulk vacuum of the scalar field is given as
\begin{eqnarray}\label{X(z)}
X(z)
=
\frac{v(z)}{2}~\mathbf{1}_2,
\end{eqnarray}
where $\mathbf{1}_2$ means a $2 \times 2$ unit matrix,
and $v(z)$ is given in Table.\ref{Tbl01}.
%---
%It is known that
$v(z)$ behaves around the boundary as
\begin{eqnarray}\label{vzero}
v(z)
&=& m_q \,\zeta\, z+\frac{\sigma}{\zeta}z^3 + {\cal O}(z^5).
\end{eqnarray}
From AdS/CFT correspondence,
$m_q$ and $\sigma$ can be interpreted
as the quark mass and quark condensate, 
respectively.
%---
As for $\zeta$, see Tables \ref{Tbl01} and \ref{Tbl02}.
%
%%%%%%%%%%%%%%%%%%%%%%%%%%%%%%%%%%%%%%%%%%
%
\begin{table}[!h]
\begin{center} \begin{tabular}{ c | l | l l l }
\hline\hline
Model & $\qquad\qquad$ $v(z)$ & \multicolumn{3}{c}{Parameters}   \\
\hline
%Ia  & $ z(A+B z^2)(1+C z^2)^{-1} $  & $A = m_q \zeta$,& \quad $B=\sigma/{\zeta}+m_q\zeta C$,            & \quad  $C=B / (\mu_d \gamma) $\\
%Ib  & $ z(A+B z^2)(1+C z^2)^{-5/4}$ & $A = m_q \zeta$,& \quad $B=\sigma/{\zeta}+\frac{5}{4}m_q\zeta C$, & \quad  $C=(B^2/ (\mu_d \gamma^2))^{2/5}$ \\
%IIa &       $ z(A+B z^2)(1+C z^4)^{-1/2}$
%    &       $ A \equiv m_q \zeta$,
%    & \quad $ B \equiv \sigma/{\zeta}$,
%    & \quad $ C \equiv (B/( \mu_d\gamma))^2$ \\
IIb &       $ z(A+B z^2)(1+C z^4)^{-5/8}$
    &       $ A \equiv m_q \zeta$,
    & \quad $ B \equiv \sigma/{\zeta}$,
    & \quad $ C \equiv (B^2/(\mu_d\gamma^2))^{4/5}$ \\
\hline\hline
\end{tabular}
\caption{
These are taken
from Ref.\cite{Sui:2009xe}.
%---
The numerical values
for the parameters appearing here are given
in Table \ref{Tbl02}, 
where these are fixed
by minimizing the breaking of the Gell-Mann-Oakes-Renner relation,
$f_{\pi}^2~m_{\pi}^2 = 2m_q~\sigma$,
at the $1\%$ level and the experimental values:
$m_\pi=139.6$ MeV and $f_\pi=92.4$ MeV.
}
\label{Tbl01}
\end{center}
\end{table}
%
%%%%%%%%%%%%%%%%%%%%%%%%%%%%%%%%%%%%%%%%%%
%
\begin{table}[!h]
\begin{center}
\begin{tabular}{c|cccccc}
\hline\hline
Model
& $\lambda$
& $m_q$ (MeV)
& $\sigma^{\frac{1}{3}}$ (MeV)
&  $\gamma$
& $\mu_g$
& $\mu$ \\
\hline
%IIa & 0 & $4.44$ & $265$ & $0.153$ & $363$ & $0.289$ \\
IIb & 0 & $4.07$ & $272$ & $0.112$ & $257$ & $0.205$ \\
%IIa & 9 & $6.95$ & $228$ & $0.30$  & $363$ & $0.289$ \\
IIb & 9 & $6.79$ & $229$ & $0.20$  & $257$ & $0.205$ \\
\hline\hline
\end{tabular}
\caption{
The numerical values
for the parameters appearing in Table \ref{Tbl01}.
These are taken from Ref.\cite{Sui:2009xe}.
%---
$\mu \equiv \mu_g/(2\pi T)$ is evaluated 
at $T=0.2$GeV, 
$\zeta=\sqrt{3}/(2\pi)$ 
and
$\mu_g = \sqrt{3}\mu_d$. 
%---
Here,  
we notice that
we will treat $\mu_g$ 
as a temperature-depending parameter.  
This way is different 
from the way 
in our previous paper~\cite{Cui:2011ag}. 
We will discuss this matter 
in this section.
%The numerical values 
%in this Table are taken from our previous paper\cite{Cui:2011ag}.
}
\label{Tbl02}
\end{center}
\end{table}

%In the last of this chapter,
Here, 
we would like to mention 
the field theory
dual to our model (\ref{action01}) 
with the geometry (\ref{metric_dads_01}).
%---
Although 
this is not an exact statement
because 
our model is a bottom-up model
and does not stand on
the configuration of the D-branes,
%---
the dual field theory 
we will assume in this paper
would be
$D=1+3$ SU$(N_c)$ gauge theory 
in large 't Hooft coupling and the large-$N_c$ limit
at finite temperature, 
which has a $U(1)$ baryon symmetry 
and $SU(N_f)_L \times SU(N_f)_R$ chiral flavor symmetry with $N_f=2$
as global symmetries.

% and no charge other than these is concerned for the global symmetries
%in the model.
%As for the supersymetry, it is uncertain.

We have shown~\cite{Cui:2011wb} that 
the quark number susceptibility 
in the model (\ref{action01}) on the geometry (\ref{metric_dads_01}) 
blows up 
when the temperature is around $160 \sim 190$ MeV,
which is considered 
to have a relation 
with the chiral symmetry breaking/restoration. 
%---
We have further examined the mass spectra 
for the vector and axial-vector mesons 
as the function of temperature~\cite{Cui:2011ag}, 
which has also been shown to blow up around $200$ MeV.
%---

The temperature and the entropy density
in the dual field theory are given
from Hawking temperature and the Bekenstein-Hawking formula as
\begin{eqnarray}\label{TH_and_s}
T_H = \frac{1}{\pi z_0}
\quad {\rm and} \quad
s = \frac{a^3(z_0)}{4G_5}=\frac{N_c^2}{2\pi}\left\{ \left(\pi T \right)^2 +\mu_g^2 \right\}^{3/2},
\end{eqnarray}
where we have used the relation \textcolor{black}{$G_5 = \pi /(2N_c^2)$}.
%---
The relations of $g_{U(1)}$ and $g_{SU(2)}$
to the soft-wall model are given
as~\cite{Karch:2006pv}
\begin{eqnarray}
\textcolor{black}{g_{U(1)}^2
=
\frac{16\pi^2}{N_c^2}}
\quad {\rm and} \quad
g_{SU(2)}^2
=
\frac{12\pi^2}{N_c}.
\end{eqnarray}
\newline

The equation for the dilaton can be obtained
from the equation of motion
for the scalar field as
\begin{eqnarray}\label{eomPhi2}
\Phi'(z)
=
\frac{3a'(z)}{a(z)}
+
\frac{\big(f(z)v'(z)\big)'}{f(z )v'(z)}
%- \frac{3}{z} +\frac{3\Omega'(z)}{\Omega(z)}
-\frac{a^2(z)}{f(z)v'(z)}
\left(
m_X^2v(z)+\frac{\lambda}{2}v^3(z)
\right).
\end{eqnarray}
It will be noticed that in our previous paper \cite{Cui:2011ag},
we  fixed $X(z)$ with considering a regularized term $ v_1(z) \ln f(z)$ as
$X(z) = \frac{1}{2}\Big\{ v_1(z) \ln f(z) + v_0(z) \Big\}~\mathbf{1}_2$
(and in accordance with this extra term, the coupling parameter $\lambda$ has to be modified),
where $v_0(z)$ in this equation corresponds to $v(z)$ in this paper.
%---
The explicit form for $v_1(z)$ is referred to in Ref. \cite{Cui:2011ag}. As it can be seen that
without considering the extra term $v_1(z) \ln f(z)$, the dilaton
$\Phi'(u)$ in the vicinity of the horizon starts
with the order $(u-1)^{-1}$ as
\begin{eqnarray}
\Phi'(u)= \frac{\Phi_{-1}(T,\mu)}{u-1}  + \mathcal{O}(1)
\end{eqnarray}
with the numerators, %for model IIa and IIb respectively as
\begin{eqnarray}
%\label{omega_IIA}
%\Phi_{-1}(T,\mu) &=&
%-\frac{1}
%{4 \left\{A \left(C(\mu)z_0^4-1\right)-B z_0^2 \left(C(\mu) z_0^4+3\right)\right\}}
%\Big[
%A
%\Big\{
%m_X^2
%\left( 4 \mu ^2 + 1 \right)
%\left( C(\mu) z_0^4 + 1 \right)
%-4 C(\mu)z_0^4 +4
%\Big\}
%\nonumber \\ &&
%+
%B z_0^2
%\Big\{
%m_X^2
%\left(4 \mu ^2+1\right)  \left(C(\mu)
%   z_0^4+1\right)+4 \left(C(\mu) z_0^4+3\right)
%\Big\}
%\Big],\\
% , \textrm{for IIa model},\\
%---
\label{omega_IIB}
\Phi_{-1}(T,\mu) &=&
-\frac{1}
{A \left(6 C(\mu) z_0^4-4\right)-2 B z_0^2 \left(C(\mu)z_0^4+6\right)}
\Big[
A
\Big\{ m_X^2 \left(4 \mu ^2+1\right) \left(C(\mu) z_0^4+1\right)-6 C(\mu) z_0^4+4\Big\}
\nonumber \\ &&
+Bz_0^2 \Big\{ m_X^2 \left(4 \mu ^2+1\right)  \left(C(\mu) z_0^4+1\right)+2 \left(C(\mu)z_0^4+6\right)\Big\}
\Big],
\end{eqnarray}
where
$u \equiv z^2/z_0^2$
and
$\mu \equiv \mu_g /(2 \pi T)$ 
[later, they are defined at Eqs.(\ref{coordu}) and (\ref{def_wqm})],
and $A$, $B$ and $C(\mu)$ are given
in Table \ref{Tbl01}.
%Here, we notice that $C(\mu)$ is a function of $\mu$ through $\mu_g$.
%---
Then, it turns out that
the contribution $\Phi_{-1}(T,\mu)/(u-1)$ appears
at the order $(u-1)^{-1}$
in the equation of motion
for the fluctuations
at the vicinity of the horizon.
%---
Therefore, 
since $\Phi_{-1}(T,\mu)/(u-1)$ is multiplied by $(u-1)^{-1}$ in the equation of motion,
finally the contribution $\Phi(\mu)$ appears
at the order $(u-1)^{-2}$
in the equation of motion.
%---
This prevents the solutions of the fluctuations
from taking the infalling boundary condition,
where the infalling boundary condition is determined
at the order $(1-u)^{-2}$
in the equations of motion.
%---
For this reason, 
it has been found in Ref.\cite{Cui:2011ag} 
introducing the extra term $v_1(z) \ln f(z)$ is useful 
in regularizing the divergence and obtaining the infalling boundary condition healthily.
%---

\begin{figure}[!h]
\begin{center}
\includegraphics[width=70mm,clip]{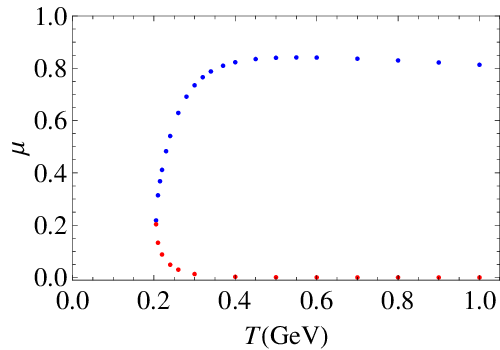}
\includegraphics[width=70mm,clip]{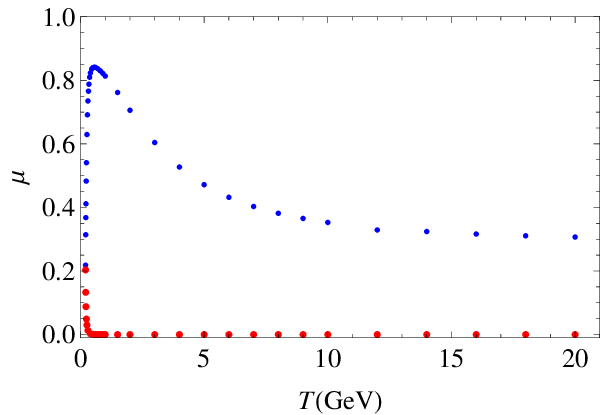}
\end{center}
\caption{
We plot
the positive and real values of $\mu$
found as the function of temperature.
%---
They are obtained 
from vanishing the numerator $\Phi_{-1}(T,\mu)$ at the order $(u-1)^{-1}$
in Eq.(\ref{omega_IIB}).
As a consequence, 
one can healthily take the infalling boundary condition.
%---
%Above two and below 
%The two figures represent the results 
%for the model IIa and IIb respectively,
%The two figures represent 
%the results 
%for the model IIb,
%and 
The left  and right plots are different just in the scale of the $x$ axis.
%---
The lower curve (red points) and upper curve (blue points) 
in each figure represent two branches of the solution
for a given temperature.
%---
}
\label{Fig_mu}
\end{figure}

However,  it turns out in this paper that 
the extra term $v_1(z) \ln f(z)$ makes the analysis 
very complicated.
Despite this, 
the reason we performed an analysis in our previous paper~\cite{Cui:2011ag} was because  
we used  numerical analyses using a shooting method. 
In this paper we will discard the way of using the extra term $v_1(z) \ln f(z)$, 
and we will take $\mu$  as a temperature-depending parameter $\mu(T)$ 
so that one can adjust the parameter $\mu$ 
to make the numerator $\Phi_{-1}(T,\mu)$ vanishes at the order $(1-u)^{-1}$. 
%However, 
%%for our present purpose in this paper, 
%it turns out in this paper that the extra term $v_1(z) \ln f(z)$ makes
%the hydrodynamic analysis very complicated.  
%Thus we have to discard
%the technique adopted in ref.\cite{Cui:2011ag} 
%%with introducing the extra term $v_1(z) \ln f(z)$ and find an alternative way. 
%A simple idea is to consider the energy scale $\mu$ (or $\mu_g$) as a temperature-depending parameter $\mu(T)$, so that one can adjust the %parameter $\mu$ to make the numerator $\Phi_{-1}(T,\mu)$ 
%\textcolor{black}{vanishes} at the order $(1-u)^{-1}$. 
%%---
%\textcolor{black}{
As a result,  
we obtain the $\mu$  
which can vanish the numerator $\Phi_{-1}(T,\mu)$ 
at each temperature  
as shown in Fig.\ref{Fig_mu}, 
and finally, 
we can take the infalling boundary condition. 
%}
%
%

In Fig.\ref{Fig_mu}, 
one can find that  
there are
two branches for the solution 
represented by
%which are plotted 
%in each figure by 
the lower curve~(red) and upper curve~(blue)
for a given temperature.
%---
Then, the question is which branch is physically meaningful.
%---
Before answering the question, 
we will make a comment
on the relation of $\mu$
in our previous paper \cite{Cui:2011ag} 
which is shown in the rightmost column of Table {\ref{Tbl02}}
and $\mu$ in this paper given in Fig.\ref{Fig_mu}.
%---

%The results represented by \textcolor{red}{the upper curve (blue points)}
%in the model IIa is constant at low temperature.
%It can be seen that the constant value
%obtained from numerical computation in this paper
%mostly matches with the value
%given in the rightmost column of Table {\ref{Tbl02}},
%though they are obtained independently.
%%---
%Next, 

For the result of model IIb, 
the branch starts at $T_c=0.2$ GeV, 
and the value of $\mu$ at the point 
at which the branch arises 
is mostly the same value of $\mu$  
in the rightmost column of Table \ref{Tbl02}  
despite that these two are obtained independently.  
%%---
Then, let us consider which branches we should take.

First, 
we recall that $\mu$ is the factor
appearing in the factor $a^2(z)$
in Eq.(\ref{metric_dads_01}) as
$
a^2(z)
=
1/z^2 + \mu_g^2
=
1/z^2 + (2\pi T \mu)^2
$.
Then, 
one can see that,  
if $\mu$ remains finite
at a high temperature, 
as the temperature increases,
our bulk space-time becomes completely different 
from the Schwarzschild $AdS_5$ black hole space-time. 
%---
It means that 
if $\mu$ does not vanish 
as the temperature goes up, 
the symmetry in the gravity side
corresponding  to the conformal symmetry 
in the dual field theory  side vanishes, 
and 
the gauge/gravity correspondence 
in our paper 
becomes invalid.
%---
%\textcolor{black}{
On the other hand, 
if $\mu$ vanishes 
at a high temperature, 
as the temperature increases, 
our bulk space-time goes back 
to the Schwarzschild $AdS_5$ blackhole space-time, 
and the symmetry 
in the gravity side can go back 
to the symmetry of $AdS_5$. 
%---
At that time, 
since our model (\ref{action01}) has the extra terms (dilaton, scalar field and gauge fields) 
other than the Einstein-Hilbert action,  
the dual field theory is not the ${\cal N}=4$ supersymmetric gauge theory  
even if the background geometry is the Schwarzschild $AdS_5$ blackhole space-time. 
%---
However, 
it would be a consistent condition 
as a holographic AdS/QCD model that 
the background geometry goes back 
to the Schwarzschild $AdS_5$ blackhole space-time 
in the high-temperature limit.  
%---
In this sense, 
eventually, 
the physically acceptable branch
would be the one represented 
by the lower curve~(red points) 
which goes to zero numerically 
in the region above around $T=0.35$ GeV.   
%and that would be inconsistent.
%}  

%On the other hand, if $\mu$ vanishes at high temperature,
%it shows that as temperature increases, our bulk space-time goes back
%to the Schwarzschild $AdS_5$ black hole space-time,
%and our dual field theory becomes the plasma phase
%at finite temperature as expected from QCD. 
%%---

%---
An interesting point is that
the effect of $\mu$ does not disappear abruptly 
but gradually disappears as the temperature increases.
%---
The effect of $\mu$ is a factor 
characterizing our AdS/QCD model,  
and it would be interesting
to study its effect more 
in the phenomenology 
in the future.
%---
Further, 
it would also be interesting that
the branch starts
at about $T=0.2$ GeV,
where the temperature $T_c=0.2$ GeV is roughly consistent 
with the critical temperature 
for the hadron/plasma transition.

%============================================
\section{Preliminaries for hydrodynamic analysis}
\label{Chap:Preliminaries}
%============================================

%----------
%\noindent
%{\bf \underline{Indies}:}\\
%----------
Before making a hydrodynamic analysis, let us first clarify the indices used in this paper 
\begin{eqnarray} \label{v1}
\textrm{$M,N$}     &=& \textrm{$x_0$, $x_1$, $x_2$, $x_3$, $z$ or } 0,~1,~2,~3,~4\\
\textrm{$\mu,\nu$} &=& \textrm{$x_0$, $x_1$, $x_2$, $x_3$ or }  0,~1,~2,~3\\
\textrm{$i,j,k$}   &=& \textrm{$x_1$, $x_2$ or } 1,~2\\
\textrm{$\alpha,\beta$}   &=& \textrm{$x_1$, $x_2$, $x_3$ or } 1,~2,~3\\
\textrm{$a,b,c$}   &=& \textrm{1~,2~,3~[indices for the $SU(2)$ algebra]}
\end{eqnarray}
where $x_0$ is coordinate of time; 
$x_1$, $x_2$ and $x_3$ are coordinates of space
on the boundary, 
and $z$ is radial direction of the bulk. % space-time.

We are going to examine the retarded Green function for
the scalar mode of the $U(1)$ baryon current, the vector mode of the $SU(2)$ flavor current and the shear viscosity
from the tensor mode of the energy-momentum tensor through AdS/CFT correspondence.
%----------
For this purpose, we will analyze the hydrodynamics
in the bulk gravity of the fluctuations,
\begin{eqnarray}
  g_{MN} &\longrightarrow&   G_{MN} = g_{MN} + h_{MN}, \label{bk_Gd} \\
  g^{MN} &\longrightarrow&   G^{MN} = g^{MN} - h^{MN}, \label{bk_Gu} \\
     A_M &\longrightarrow&   B_{0M} = 0 + A_M        , \label{bk_A}  \\
   B^a_M &\longrightarrow& B^a_{0M} = 0 + B^a_M      , \label{bk_B}
\end{eqnarray}
where the background of the gauge fields is vanishing.
Without confusing, we simply use the same character
for the total gauge fields and the fluctuation of gauge fields.
%---
These fluctuations are in linear order in the equations of motion.
%---
In this paper, we will treat the dilaton as the static background.
\newline

We show  the classification of fluctuations in the $SO(2)$ little group
for the $x$ and $y$ directions as
\begin{eqnarray}
&& \mbox{Scalar mode: } h_{00},\, h_{03},\, h_{33},\, h_{zz},\, h_{0z},\, h_{3z},\, h^k_{~k} \quad {\rm and} \quad A_0,\, A_3,\, A_z,  \nonumber \\
&& \mbox{Vector mode: } h_{0i},\, h_{3i},\, h_{zi} \quad {\rm and} \quad A_i,
\label{FlucClass}
\\
&& \mbox{Tensor mode: } h_{ij}-\delta_{ij} h^k_{~k}/2.\nonumber
\end{eqnarray}
One may see the correspondence of the above classification
with the one given in Sec.\ref{Chap:Hydro}.
%-----

In AdS/CFT correspondence, the fluctuations of the bulk gauge field $A_\mu$ act
as the source for the $R$-charge current $J_\mu$
on the dual field theory.
%-----
Since it is known that 
the $U(1)$ baryon number 
can be regarded 
as an analog of $R$ charge in AdS/CFT correspondence,
we will consider that 
$A_\mu$ acts as the source
for the global $U(1)$ baryon charge current $J_\mu$
~(For example, see Ref.\cite{RchargeBcharge}).
%-----
Further, $B_{L,\mu}$ and $B_{R,\mu}$ also act
as the source for the global $SU(N_f)_L \times SU(N_f)_R$ chiral flavor currents
%$\bar{q}_L \gamma^\mu t^a q_L$ and $\bar{q}_R \gamma^\mu t^a q_R$,
in which $N_f=2$ in this paper.
%-----
The gravitational perturbations in the bulk
act as the source for the stress-energy tensor $T_{\mu\nu}$
on the dual theory.

In our following analysis, for convenience, 
we will define the normalized radial coordinate
\begin{eqnarray}\label{coordu}
u \equiv %\frac{r_0^2}{r^2}
\frac{z^2}{z_0^2},
\end{eqnarray}
and the normalized frequency, momentum and the factor
\begin{eqnarray}\label{def_wqm}
\nw \equiv \frac{\omega}{2\pi T},\quad
\nq \equiv \frac{q}{2\pi T}
\quad {\rm and} \quad
\nmu \equiv \frac{\mu_g}{2\pi T}.
\end{eqnarray}

%
%$\partial^\mu A_{L,\mu}=\partial^\mu A_{R,\mu}=0$
%to get rid of unphysical polarization.

Finally, let us mention the gauge fixing condition in our present consideration.
%-----
Like other papers~(for example, Ref.\cite{Misumi}),
we choose the axial gauge condition and the Landau gauge as follows
\begin{eqnarray}
A_z&=&0,
\quad
B^a_z=0
\quad \textrm{and} \quad
h_{Mz}=0, \\
\partial^\mu A_{\mu}&=&0
\quad \textrm{and} \quad
\partial^\mu B_{L,\mu}=\partial^\mu B_{R,\mu}=0,
\end{eqnarray}
where the Landau gauge is imposed in the four-dimensional space-time
on the boundary.
%where ``$*$'' means any indices except for the radial direction.
%\newline

%====================================================
\section{Analysis on the $U(1)$ baryon current}
\label{Chap:U(1)_baryon_current}
%====================================================

%
%\noindent
%{\bf \underline{Equations of motion:}}\\
In this section, we will first perform a hydrodynamic analysis
in the scalar mode shown in Table.(\ref{FlucClass}) of the $U(1)$ baryon current.
%---
To begin with,
we can write
the equation of motion
for the fluctuation of $A_M$ in terms of the field strength as
\begin{eqnarray}\label{eq:maxwell}
\partial_M \left(\sqrt{-g(u)}e^{-\Phi(u)} F^{MN}(u) \right) = 0,
\end{eqnarray}
where $g(u) \equiv \det g_{M N}(u)$.
We will carry out the analysis in the momentum space by performing the plane wave expansion as
\begin{eqnarray}\label{eq:fourier}
A_{\mu}(u, x_0, x_3)
=
\int\! \frac{dw\, dq}{(2\pi)^2}\,
e^{-i \omega x_0 + i q x_3}
A_{\mu} (u, \omega, q)\,,
\end{eqnarray}
where we have rotated the direction of momentum to $x_3$ 
and paid attention to the same notation used
in the gauge fields before and after Fourier transformation:
\begin{eqnarray}
& & \omega g^{00}(u) A_0'(u) - q g^{33}(u) A_3'(u) = 0~,
\label{eq:maxwell1} \\
& &
\partial_u \left( e^{-\Phi(u)} \sqrt{-g(u)} g^{00}(u) g^{uu}(u) A_0'(u) \right)
 =  e^{-\Phi(u)}\sqrt{-g(u)} g^{00}(u) g^{33}(u) \left( \omega q A_3(u) + q^2 A_0(u) \right),
\label{eq:maxwell2} \\
& &
\partial_u \left( e^{-\Phi(u)} \sqrt{-g(u)} g^{33}(u) g^{uu}(u) A_3'(u) \right)
 =  e^{-\Phi(u)}\sqrt{-g(u)} g^{00}(u) g^{33}(u) \left( \omega q A_0(u) + w^2 A_3(u) \right),
 \label{eq:maxwell3} \\
& &
\partial_u \left( e^{-\Phi(u)}\sqrt{-g(u)} g^{\alpha \alpha}(u) g^{uu}(u) A'_\alpha(u) \right)
 = e^{-\Phi(u)}\sqrt{-g(u)}g^{\alpha \alpha}(u) \left(g^{00} \omega^2 A_\alpha(u) + g^{33} q^2 A_\alpha(u) \right),
\label{eq:maxwell4}
\end{eqnarray}
where $~{}' \equiv \partial_u$.

From Eqs.(\ref{eq:maxwell1}) and (\ref{eq:maxwell2}), the equation of motion
for $A'_t$ can be obtained as follows: 
\begin{eqnarray}
\left( \frac{z_0}{2}\right)^2
\frac{d}{du}
\left[
\frac
{\partial_u ( e^{-\Phi(u)} \sqrt{-g(u)} g^{00}(u) g^{uu}(u) A_0'(u))}
{e^{-\Phi(u)} \sqrt{-g(u)} g^{00}(u) g^{33}(u)}
\right]
-
\left(\frac{g^{00}(u)}{g^{33}(u)} \nw^2 + \nq^2\right) A_0'(u) =0.
\label{eq:master1}
\end{eqnarray}
For a technical reason, we will perform a rescaling as
\begin{eqnarray}
-\frac{2}{z_0^2}
e^{-\Phi(u)} \sqrt{-g(u)} g^{00}(u) g^{uu}(u) A_0'(u) \equiv {\cal A}_0'(u),
\end{eqnarray}
where we attach the factor $-2/z_0^2$ 
in the front 
to make the leading term of the rescaling
as a unit in the expansion around $u=0$. 
Note that when taking $\mu=0$, 
it recovers the case of the Schwarzschild $AdS_5$ black hole geometry.
%---
With the above rescaling, the equation of motion is given by 
\begin{eqnarray}\label{eq:master1}
\left( \frac{z_0}{2}\right)^2
\frac{d}{du}
\left[
\frac{\partial_u {\cal A}_0'(u)}{e^{-\Phi(u)} \sqrt{-g(u)} g^{00}(u) g^{33}(u)}
\right]
=
\frac{\frac{g^{00}(u)}{g^{33}(u)} \nw^2 + \nq^2}{e^{-\Phi(u)} \sqrt{-g(u)} g^{00}(u) g^{uu}(u)} {\cal A}_0'(u).
\end{eqnarray}
%where the factor $(z_0/2)^2$ is attached to change $\omega$ and $k$ to $\nw$ and $\nq$.

To make the hydrodynamic analysis, 
let us formally write down the solution
to the first-order hydrodynamics as follows: 
\begin{eqnarray}\label{eq:At01}
A_0'(u) &=&
%\textcolor{red}{-\frac{z_o^2}{2}}
C^{(0)}_0(u)
\frac{e^{\Phi (u)}}{\sqrt{1+4\mu^2 u}}
(1-u)^{-i \nw/2}
\left(
1+ \nw F^{(0)}_1(u) + \nq^2 G^{(0)}_1(u) + \cdots
\right),
\end{eqnarray}
where $C^{(0)}_0(u)$, $F^{(0)}_1(u)$ and $G^{(0)}_1(u)$ are going to be fixed below.
%---
It is noticed that 
the infalling boundary condition has been used 
to yield the factor $(1-u)^{-i \nw/2}$
in the solution.
%---
This is realized 
by taking the special solution of $\mu$, 
as shown in Fig.\ref{Fig_mu}, 
with requiring
the numerator in the contribution at the order of $(u-1)^{-1}$ to be vanishing in the dilaton $\Phi'(u)$.
%---
Then, we can see readily that $C^{(0)}_0$ can be treated as a constant,
\begin{eqnarray}
C^{(0)}_0(u) \equiv C^{(0)}_0,
\end{eqnarray}
and the rest coefficients are obtained
by the following equations:
\begin{eqnarray}
\label{eq:F1}
\frac{z_0^2}{8}
\partial_u
\left\{
\frac{
i C^{(0)}_0
+
2(1-u) \partial_u F_1^{(0)}(u)
}
{
(1-u)e^{-\Phi(u)} \sqrt{-g(u)} g^{00}(u) g^{33}(u)
}
\right\}
&=&0,\\
%-----
\label{eq:G1}
\frac{z_0^2}{4}
\partial_u
\left(
\frac{
\partial_u
G_1^{(0)}(u)
}
{
e^{-\Phi(u)} \sqrt{-g(u)} g^{00}(u) g^{33}(u)
}
\right)
&=&
\frac{C^{(0)}_0}
{
e^{-\Phi(u)} \sqrt{-g(u)} g^{uu}(u) g^{00}(u)
}
,
\end{eqnarray}
where the former and latter can be obtained 
from the order at $\nw$ and $\nq^2$
in Eq.(\ref{eq:At01}). 
Let us solve the above equations %order by order
by taking the following three steps:
\begin{enumerate}
\item
performing the integration on the whole equations
to get rid of the derivative $\partial_u\{ \cdots \}$ or $\partial_u( \cdots )$,
\item
rewriting the equations in the form ``$\partial_u F^{(0)}_1 = \cdots$'' or ``$\partial_u G^{(0)}_1 = \cdots$'',
\item
carrying out the integration over the whole equations
to obtain $F^{(0)}_1$ and $G^{(0)}_1$.
\end{enumerate}
We then obtain the coefficients as follows
\begin{eqnarray}
F^{(0)}_1(u)
&=&
C^{(0)}_0
\left(
F_B^{(0)}
-
\frac{2 F_H^{(0)} e^{-\Phi (0)} \log (u)}{z_0^2}
\right)
-
C^{(0)}_0
\left(
\frac{i}{2}
-
\frac{2  F_H^{(0)} e^{-\Phi (0)} \left(\Phi '(u)|_{u=0} -2 \mu ^2\right)}{z_0^2}
\right)
u
+ {\cal O}(u^2),
\label{eq:U1F}
\nonumber\\
\\
%-----
G^{(0)}_1(u)
&=&
C^{(0)}_0
\left(
G_B^{(0)}
-
\frac{2 F_H^{(0)} e^{-\Phi (0)} \log (u)}{z_0^2}
\right)
+
C^{(0)}_0
\left(
1
+
\frac{2  G_H^{(0)} e^{-\Phi (0)} \left(\Phi '(u)|_{u=0}-2 \mu ^2\right)}{z_0^2}
\right)
u
+ {\cal O}(u^2)
\nonumber\\
\label{eq:U1G}
\end{eqnarray}
with
\begin{eqnarray}
C^{(0)}_0 = -\frac{\nq z_0^2(A_0^{(0)} \nq + A_3^{(0)} \nw)}{2(G_H \nq^2 + F_H\nw)}.
\end{eqnarray}
Here $C^{(0)}_0$ has been fixed
by using Eq.(\ref{eq:maxwell2})
with the Dirichlet boundary condition that
the boundary value of $A_0$ and $A_3$ are given
by $A^{(0)}_0$ and $A^{(0)}_3$. 
Note that if the full integration in step $1$ or $3$ is difficult, 
instead of it,
one can first carry out the Tyler expansion around $u=0$, 
and after that, 
perform integration
toward its low-order terms.
%---
As we will use eventually the GKP-W relation, 
the resulting solutions are the ones
around the boundary, 
%---
where $F_H^{(0)}$, $F_B^{(0)}$, $G_H^{(0)}$ and $G_B^{(0)}$ are the integral constants.
%---
$F_H^{(0)}$ and $G_H^{(0)}$ appear
from the integration in step $3$,
and they can be taken arbitrarily.
%---
On the other hand, $F_B^{(0)}$ and $G_B^{(0)}$ appear
from the integration in step $1$, and they are, in general, 
fixed such that the solutions do not diverge at $u=1$, 
%---
where the subscripts ``$H$'' and  ``$B$'' denote the relevant constants which are fixed
from the horizon and the boundary, respectively.
%---

To put it more concretely, after step $2$, one can see that
$F_1^{(0)}{}'(u)$ and $G_1^{(0)}{}'(u)$ behave  as
$\displaystyle F_1^{(0)}{}'(u) ~{\rm}, G_1^{(0)}{}'(u) \sim 1/(u-1)$ around $u=1$, 
which generally leads to the solution with logarithmic divergence, such a solution is ill defined.
To obtain a physically meaningful solution, 
a simple way is to vanish these contributions 
by exploiting the integral constants $F_H^{(0)}$ and $G_H^{(0)}$ 
so that the numerators become zero.
%---

In the actual calculation, $F^{(0)}_H$ can easily be fixed  to be
\begin{eqnarray}
F^{(0)}_H=-\frac{i C^{(t)}_0 e^{\Phi(1)} z_0^2}{2\sqrt{1+\mu^2}}, 
\end{eqnarray}
while fixing $G_H$ is, in general, 
complicated as the integrating becomes difficult due to the dilaton.
For this reason, we have to perform integration around $u=1$ to fix the integral constant $G_H$.
It will be shown that, as long as the integral constant $G_H$ is chosen appropriately, 
we can arrive at the needed solution which has no logarithmic divergence  and becomes well defined 
in the whole bulk. Let us conduct step $1$
with an expansion around $u=1$:  
\begin{eqnarray}
&&
\hspace{-4mm}
\frac{\partial_u G^{(0)}_1(u)}{e^{-\Phi(u)} \sqrt{-g(u)} g^{00}(u) g^{33}(u)}
\nonumber\\
&&
\hspace{-8mm}
=
\int du \frac{C^{(0)}_0 }
{
e^{-\Phi(u)} \sqrt{-g(u)} g^{uu}(u) g^{00}(u)
}
\nonumber\\
&&
\hspace{-8mm}
=
%G_H +
\int du
\Big\{
C_0 + C_1(u-1)  + C_2(u-1)^2  + C_3(u-1)^3 + \cdots
\Big\}
\nonumber\\
&&
\hspace{-8mm}
=
G^{(0)}_H
+ C_0 u
+ C_1 \left( \frac{u^2}{2} - u \right)
+ C_2 \left( \frac{u^3}{3} - u^2 + u \right)
+ C_3 \left( \frac{u^4}{4} - u^3 + \frac{3}{2}u^2 - u \right) + \cdots
\nonumber\\
&&
\hspace{-8mm}
=
G^{(0)}_H +
\sum_{n=0}^\infty \frac{(-1)^n}{n+1}C_n
%\left(
%C_0
%- \frac{C_1}{2}
%+ \frac{C_2}{3}
%- \frac{C_3}{4}
%+ \frac{C_3}{5} + \cdots
%\right)
+ C_0 t
+ C_1 \left( \frac{t^2}{2} - t \right)
+ C_2 \left( \frac{t^3}{3} - t^2 + t \right) %\nonumber \\ &&
+ C_3 \left( \frac{t^4}{4} - t^3 +  \frac{t^2}{2} - t \right) + \cdots
\end{eqnarray}
with
\begin{eqnarray}
C_n \equiv
%\sum_{n=1}^\infty
\frac{C^{(0)}_0 }{n!} \,
\partial_u^{(n)}
\left(
\frac{1}
{
e^{-\Phi(u)} \sqrt{-g(u)} g^{uu}(u) g^{00}(u)
}
\right)\Bigg|_{u=1} \quad {\rm and} \quad t\equiv u-1,
\end{eqnarray}
where ``$(n)$'' means the number of the derivative with regard to $u$.
%---
In the above equations,
we have performed an expansion
from the second to the third  lines and written its result in a symbolic way.
%---
From the third to the fourth lines,
we have explicitly written the integral constant $G^{(0)}_H$.
%---
From the fourth  to the fifth lines,
we have changed the variable $u$ to $t \equiv u-1$.
%---
It is seen that, 
when we rewrite it in the form that ``$\partial_u G^{(0)}_1= \cdots$'', 
there is no term with $1/(u-1)$ in the expansion beyond the constant term.
%---
Therefore, we can readily obtain the expression of $G^{(0)}_H$ to be
\begin{eqnarray}\label{G_H_U(1)}
G^{(0)}_H
= - \sum_{n=0}^\infty
\frac{(-1)^{n}}{n+1}C_n
\equiv
\frac{ C^{(0)}_0 z_0^2 }{2 e^{-\Phi(1)} \sqrt{1+4\mu^2}}G_A\big(\mu,\big\{\Phi^{(n)}\big\}\big),
\end{eqnarray}
where we have rewritten symbolically the concise expression for convenience
in the following actual calculation.
%---
Here, 
$
G_A\left(\mu,\{\Phi^{(n)}\} \right)
$
is a function with
$
\big\{\Phi^{(n)}\big\}
\equiv
\big\{
%\Phi(u=1),
\Phi^{(1)}(u)|_{u=1},
\Phi^{(2)}(u)|_{u=1},
\Phi^{(3)}(u)|_{u=1},\cdots
\big\}
$
, and when $\mu=0$ and $\big\{\Phi^{(n)}\big\}=\big\{0\big\}$,
one has  $G_A(0,\{0\})=1$. 
In this paper, $G^{(0)}_H$ will be treated symbolically.
\newline

It can be shown that 
the solutions near the boundary behave as follows: 
\begin{eqnarray}
A_0'(u) &=&
-\frac{2 C^{(0)}_0}{z_0^2}
\left(G^{(0)}_B \nq^2 + F^{(0)}_B \nw \right) \log(u)
+C^{(0)} e^{\Phi(0)}\left(1+G^{(0)}_B \nq^2 +F^{(0)}_B \nw \right),\\
%====
A_3'(u) &=&
\frac{2 C^{(0)}_0}{z_0^2}
\nw
\left(G^{(0)}_B \nq + F^{(0)}_B \frac{\nw}{\nq}\right)\log(u)
-C^{(0)}_0 e^{-\Phi(0)} \frac{\nw}{\nq}
\left( 1 + G^{(0)}_B \nq^2 +F^{(0)}_H \nw \right).
%%====
%A_i(u) &=&
%A_i^0\left(1 -  \nw \frac{i}{4} \log (2) + i u \nw e^{-\Phi (1)} \sqrt{1+\mu^2} \right),
\end{eqnarray}

%-----

The boundary action $S_0$ is given
from the on-shell action in the quadratic order of fluctuation as
\begin{eqnarray}
S_0=
\frac{1}{g^2 z_0^2}
\int \frac{dw\, dq}{(2\pi)^2}
\left\{
\left(1+\frac{4\mu ^2 \sqrt{u}}{z_0} \right) A_0(u) A_0'(u)
-
\left(1+\frac{4\mu ^2 \sqrt{u}}{z_0} \right)
%\left(
\sum_{\alpha} A_\alpha(u) A_\alpha'(u)
%A_1(u) A_1'(u)
%+
%A_2(u) A_2'(u)
%+
%A_3(u) A_3'(u)
%\right)
\right\}.
\end{eqnarray}
By using the prescription 
for obtaining the retarded Green function
in AdS/CFT correspondence~\cite{Son:2002sd}, 
we arrive at
the two-point retarded Green function of the $U(1)$ baryon charge current
in the scalar mode as 
\begin{eqnarray}
G_{\rm (R)}^{00}
&=&
\frac{N^2 T^2}{8}
\frac
{ e^{\Phi (0)-\Phi (1)} \sqrt{4 \mu ^2+1} k^2}
{2 i \pi T \omega -  G_0(\mu,\{\Phi^{(n)}\}) k^2 }
,\\
%=====
G_{\rm (R)}^{33}
&=&
\frac{N^2 T^2}{8}
\frac
{ e^{\Phi (0)-\Phi (1)} \sqrt{4 \mu ^2+1} \omega^2}
{2 i \pi T \omega -  G_0(\mu,\{\Phi^{(n)}\}) k^2 }.
\end{eqnarray}
From these results, we can read out  the diffusion constant (\ref{difconst}) as
\begin{eqnarray}\label{U(1)dc}
D
=
\frac{G_A(\mu,\{\Phi^{(n)}\})}{2\pi T}.
\end{eqnarray}

%====================================================
\section{Analysis on the $SU(2)$ vector and axial-vector flavor currents}
\label{Chap:SU(2)_flavor_current}
%====================================================

In this section, 
we will perform the hydrodynamic analysis
for the vector mode 
shown %as the $SU(2)$ vector current and axial-vector flavor current of 
in Table (\ref{FlucClass}).
%---
To begin with, let us combine the $SU(2)$ gauge fields
into the vector field $V_M$ and the axial-vector field $W_M$, 
\begin{eqnarray}\label{def_VW}
V_M(u) \equiv \frac{1}{2}(B_{L,M}(u)+B_{R,M}(u))
\quad {\rm and} \quad
W_M(u) \equiv \frac{1}{2}(B_{L,M}(u)-B_{R,M}(u)).
\end{eqnarray}
%---
The equations of motion for the vector field $V^a_x$ and the axial-vector field $W^a_x$ are given as
\begin{eqnarray}
\label{eq:eomV1}
\textrm{V} \, &:& \,
D_{V,M}
\left( \sqrt{-g(u)} e^{-\Phi(u)} \, F^{a,MN}_V(u)  \right)=0,\\
%---
\label{eq:eomAV1}
\textrm{AV} \, &:& \,
D_{W,M} \left(  \sqrt{-g(u)} e^{-\Phi(u)}  \, F^{a,MN}_W(u)  \right)
+~
g_{SU(2)}^2
 e^{-\Phi(u)} \sqrt{-g(u)} ~ v^2(u)  g^{MN}(u) W^a_M(u)=0,
\end{eqnarray}
where $v(u)$ has been defined in Eq.(\ref{X(z)}),
and ``V'' and ``AV'' mean the field strength
consisting of the vector and the axial-vector fields, respectively.
$F^a_{V,MN}(u)$ and $D_{V,M}(u)$ are dictated as
$
F^a_{V,MN}(u)
\equiv
\partial_M V^a_N(u)
- \partial_N V^a_M(u)
+ i f^{abc} V_M^b(u)  V_N^c(u)
$
[$f^{abc}$ is the $SU(2)$ structure constant],
and $D_{V,M} \equiv \partial_M(u) -iV_M(u)$
~[$F_{W,MN}(u)$ and $D_{W,M}(u)$ are similar].
%---
It is seen that
Eq.(\ref{eq:eomV1}) can be obtained
from Eq.(\ref{eq:eomAV1})
by dropping the term
proportional to $g_{SU(2)}^2$.
%---
Thus, 
our main task here will focus
on solving Eq.(\ref{eq:eomAV1}).
\newline

Now let us write down the equation of motion (\ref{eq:eomAV1})
to the linear order, and its result is given by
\begin{eqnarray}\label{eq:W1}
\partial_u
\big(
e^{-\Phi(u)}\sqrt{-g(u)}g^{\alpha \alpha}(u)g^{uu}(u)W^a_\alpha{}'(u)
\big)
=
e^{-\Phi(u)}g^{\alpha \alpha}(u)
\big(
g^{00}(u)\omega^2+g^{33}(u)k^2-g_{SU(2)}^2 v^2(u)
\big)
W^a_\alpha(u).
\nonumber \\ %=0.
\end{eqnarray}
which cannot simply be solved by the double integral dictated above Eq.(\ref{eq:U1F})
due to the extra term proportional to $g_{SU(2)}^2 v^2(u)$ in the potential part.
Thus, we may factorize it as follows: 
\begin{eqnarray}
W^a_\alpha(u) \rightarrow \rho(u) {\cal W}^a_\alpha(u).
\end{eqnarray}
Then, Eq.(\ref{eq:W1}) becomes the following form
\begin{eqnarray}
\partial_u
\Big(
e^{-\Phi(u)} \sqrt{-g(u)} g^{\alpha \alpha}(u) g^{uu}(u) {\cal W}^a_\alpha{}'(u)
\Big)
+
\rho(u)
\left\{
\Omega(u)
-  e^{-\Phi(u)} \sqrt{-g} \rho g^{\alpha \alpha} (u)
(g^{00} \omega^2 +g^{33}k^2)
\right\}=0
\nonumber \\
\end{eqnarray}
with
\begin{eqnarray} \label{Omega}
\Omega(u)
\equiv
\partial_u
\Big(
e^{-\Phi(u)} \sqrt{-g(u)} g^{\alpha \alpha}(u) g^{uu}(u) \rho'(u)
\Big)
-
g_{SU(2)}^2 e^{-\Phi(u)} \sqrt{-g(u)} \rho(u) g^{\alpha \alpha}(u)  v^2(u).
\end{eqnarray}
which indicates that when $\rho(u)$ satisfies $\Omega(u) =0$,
it becomes available to obtain the solution $W_\alpha^a(u)$ from the double integral. 
Nevertheless, to fully solve the equation $\Omega(u)=0$ is technically difficult. 
Practically, considering that the needed solution at the last stage is the one
only in the vicinity of the horizon and the boundary, we shall obtain the solutions
in the expansion around $u=0$ and $u=1$.

First, let us obtain the solution in the expansion around $u=0$.
For this purpose, expanding $\Omega(u)$ given in Eq.(\ref{Omega}),
we arrive at the following form 
\begin{eqnarray}\label{eq:rho}
\big( \alpha_0 + {\cal O}(u) \big) \rho''(u)
+
\big( \beta_0 + {\cal O}(u) \big) \rho'(u)
+
\left( \frac{\gamma_0}{u}+\gamma_1+ {\cal O}(u) \right) \rho(u) = 0
\end{eqnarray}
with
\begin{eqnarray}\label{eq_expanded:rho}
\alpha_0 \equiv \frac{2}{z_0^2},
\quad
\beta_0 \equiv -\frac{2}{z_0^2}
\left(\Phi'(u)\big|_{u=0}-2\mu^2\right),
\quad
\gamma_0 \equiv \frac{1}{2}g_{SU(2)}^2 A^2,
\quad
\gamma_1 \equiv g_{SU(2)}^2 A^2
\left(B z_0^2+3 A \mu^2\right). %\nonumber\\
\end{eqnarray}
%which are in common for model IIa and IIb.
As a result, we obtain the general solution $\rho(u)$
as a linear combination of two confluent hypergeometric functions  ${}_1F_1$ and $U$
with two integral constants.
%---
We set  two integral constants in such a way that ${}_1F_1$ vanishes, 
and the leading of $\rho(u)$ in the expansion of $u=0$ becomes $1$.
%---
Eventually, we obtain $\rho(u)$ satisfying Eq.(\ref{eq_expanded:rho})
to the expanded order as
\begin{eqnarray}\label{rho_AV}
\rho(u)
=
\Exp
\left(
-\frac{\Delta+\beta_0}{2 \alpha_0}u
\right)
\Gamma
\left(
1-\frac{\gamma_0}{\Delta}
\right) U
\left(
-\frac{\gamma_0}{\Delta},
0,
\frac{ \Delta}{\alpha_0}u
\right)
\end{eqnarray}
with $\Delta \equiv \sqrt{\left(\beta_0\right)^2-4 \alpha_0 \gamma_1}$, 
where the functions $\Gamma(u)$ and $U(u)$ denote the gamma function 
and the confluent hypergeometric function, respectively.

Now we shall try to solve $\Omega(u)=0$ in the vicinity of $u=1$.
It turns out that a general solution is given by a linear coupling of
the confluent hypergeometric function $U$ and
Laguerre polynomials $L_n$. 
% with some $n$
%(The actual expression of $n$ is lengthy, and we skip to show it here).
%---
However, it is difficult to fix the integral constants analytically,
such that the constants in the solution at $u=1$ become common with
the ones at $u=0$ of Eq.(\ref{rho_AV}), 
due to the lack of information for the solution of $\rho(u)$
in the interior domain of the bulk.
Therefore, we will only treat the solution $\rho(u)$ around $u=1$ symbolically.

Having obtained $\rho(u)$ satisfying $\Omega(u)=0$ as in Eq.(\ref{rho_AV}),
we can solve the equation of motion by using the double integral similar to the previous chapter.
With the results of $\mu$ given in Table \ref{Fig_mu},
we can formally write down the solution of ${\cal W}^a_\alpha$ in the hydrodynamic expansion
with regard to $\nw$ and $\nq^2$ as
\begin{eqnarray}
{\cal W}^a_\alpha
=
C^{(a)}_\alpha (u)
(1-u)^{-i \nw/2}
\left(
1+ \nw F^{(a)}_1(u) + \nq^2 G^{(a)}_1(u) + \cdots
\right).
\end{eqnarray}
With the same way stated above Eq.(\ref{eq:U1F}),
the coefficients can be written as
\begin{eqnarray}
C^{(a)}_\alpha(u)
=
\frac
{{\cal W}_\alpha^{a}{}^{(0)}}
{1+G_B^{(a)} q^2 +F_B^{(a)} \omega} 
\end{eqnarray}
and
\begin{eqnarray}
F^{(a)}_1(u)
&=&
F_B^{(a)}
+ \left(\frac{2 F^{(a)}_H e^{\Phi (0)}}{\rho (0)^2}-\frac{i C^{(a)}_0}{2}\right) u
+ {\cal O}(u^2),\\
G^{(a)}_1(u)
&=&
G^{(a)}_B
+ \left\{ C^{(a)}_0 (\log (u)-1)+\frac{2 G^{(a)}_H e^{\Phi (0)}}{\rho^2(0)}\right\} u
+ {\cal O}(u^2),
\end{eqnarray}
where ${\cal W}_\alpha^{a}{}^{(0)}$ denotes the boundary value of ${\cal W}_\alpha^{a}$,
and $F^{(a)}_B$, $G^{(a)}_B$ and $F^{(a)}_H$, $G^{(a)}_H$ mean integral constants
which are fixed at the boundary and the horizon as the same as the integral constants 
that appeared in Eqs.(\ref{eq:U1F}) and (\ref{eq:U1G}).
%---
To be explicit, they are fixed as
\begin{eqnarray}\label{F_B_G_B_SU(2)}
F^{(a)}_H &=& \frac{1}{2} i C^{(a)}_0 \sqrt{4 \mu ^2+1} \, \rho^2 (1) e^{-\Phi (1)}, \\
G^{(a)}_H &=& -\frac{1}{2} C^{(a)}_0  \sqrt{4 \mu ^2+1} \, e^{-\Phi (1)} G_W(\mu,\{ \Phi^{(n)} \}),
\end{eqnarray}
%-----
where $G_W(\mu,\{ \Phi^{(n)} \})$ is a function as the same as Eq.(\ref{G_H_U(1)}) and $G_W(0,\{0\})=1$.
%-----
From the above analysis, we can write down the solutions
in the vicinity of $u=0$ with  taking the factor $\rho$
\begin{eqnarray}
W^a_\alpha &=&
{\cal W}_\alpha^{a}{}^{(0)}
\rho(0)
+
{\cal W}_\alpha^{a}{}^{(0)}
\left\{
\frac{4  \rho(0) e^{\Phi (0)} \left(F^{(a)}_H \omega +G^{(a)}_H q^2 \right)
+  \rho^2(0) \left(2 q^2 \log (u)-2 q^2-i \omega \right)}
{2 \rho(0) \left(F^{(a)}_B \omega +G^{(a)}_B q^2+1 \right)}
%-----------
\right.
\nonumber \\ && \left. \qquad \qquad \qquad \qquad
%-----------
+
\left(
\rho(1)+\frac{i \rho(0)  }{2} \omega
\right)
\right\}u
+{\cal O}(u^2).
\end{eqnarray}
It is noticed that $V^a_\alpha$ can simply be read from
the solution of $W^a_\alpha$  by taking $g_{SU(2)}^2= 0$.
\newline

Finally, we can obtain the two-point retarded Green function
in the dual field theory 
through the GKP-W relation. 
For this purpose,
the boundary action 
in quadratic order 
at $u=0$ is needed.
One can see that 
it is given as
\begin{eqnarray}
S_0 = -\frac{2 }{g_{SU(2)}^2 z_0^2}\left(1+\frac{4\mu ^2 \sqrt{u}}{z_0}\right)
\int \frac{dw\, dq}{(2\pi)^2}
\sum_{\alpha} W_\alpha^a(u) W_\alpha^a{}'(u).
\end{eqnarray}
which enables us to obtain the two-point retarded Green function
in the vector mode as
\begin{eqnarray}
G^{\alpha \alpha}_{\rm AV,R}
&=& \frac{i N^2 T^2}{8}
\rho(0)
\left\{
\frac{4  \rho(0) e^{\Phi (0)} \left(F^{(a)}_H \omega +G^{(a)}_H q^2 \right)
+  \rho(0)^2 \left(2 q^2 \log (u)-2 q^2-i \omega \right)}
{2 \rho(0) \left(F^{(a)}_B \omega +G^{(a)}_B q^2+1 \right)}
+
\left(
\rho(1)+\frac{i \rho(0) \omega }{2}
\right)
\right\}
.\nonumber\\
\end{eqnarray}
The two-point retarded Green function $G^{\alpha \alpha}_{\rm V,R}(u)$
for $V^a_\alpha(u)$ can simply be resulted
from the $G^{\alpha \alpha}_{\rm AV,R}(u)$
given above with vanishing $g_{SU(2)}^2$.

%============================================
\section{Analysis on the tensor mode}
\label{Chap:tensor}
%============================================

In this section, 
we will carry out 
the analysis of the tensor mode of the fluctuation 
in the dual gravity
%We are now going to carry out the analysis for the tensor mode of the fluctuation
shown in Table (\ref{FlucClass}). 
%---
Then, from these results, 
we will calculate the shear viscosity (\ref{Kuboeta})
and its ratio to the entropy density 
in the dual field theory.
%---
For this purpose, 
we will start 
with the equation of motion
for the bulk gravity, 
which can be generally given as
\begin{eqnarray} \label{Einstein_eom}
R_{MN}
-
\frac{1}{2}G_{MN} R
+
\Lambda G_{MN}
=
\kappa^2 T_{MN},
\end{eqnarray}
where $\kappa^2 = 8\pi G_5$
[$G_5$ is given below Eq.(\ref{TH_and_s})],
$\Lambda  = -6/l^2$
($l$ denotes the curvature radius of the AdS space and is taken to be 1), 
and $T_{MN}$ means the energy-momentum tensor.
%---
From Eq.(\ref{metric_dads_01}), it is given as
\begin{eqnarray}
T_{MN} \equiv
g_{MN} {\cal L}
+
g^{PQ}
\left\{
\frac{1}{g^2_{U(1)}}
F_{PM} F_{QN}
+
\frac{1}{g^2_{SU(2)}}
{\rm Tr}
\Big(
F_{L,PM} F_{L,QN}
+
F_{R,PM} F_{R,QN}
\Big)
\right\}
-2{\rm Tr}D_M X D_N X
,\nonumber\\
\end{eqnarray}
where ${\cal L}$ means the Lagranigian (\ref{action01})
except for the Einstein-Hilbert part.
%---
It is noticed that, 
when substituting the background (\ref{metric_dads_01})
into the above equation of motion for gravity, there is a deviation
proportional to $\mu_g$, which indicates that
the background (\ref{metric_dads_01}) is not an exact 
solution of the equation of motion for gravity 
as the backreacted geometry is not considered here.  
%---
Nevertheless, the effect from the backreacted geometry was found to be small\cite{BRG}, 
and also it turns out that the deviated part appears
only in the diagonal part of the equation of motion,
while in our present treatment it involves only the
off-diagonal part of the equation of motion, the $(x,y)$ component, 
so the deviated part will be discarded in this paper.

To evaluate the viscosity, 
we will consider the $(x,y)$@component
in Eq.(\ref{Einstein_eom}).
%---
Then, the energy-momentum tensor
at the linear order becomes as
$
T_{xy}
=
\Tr
\left(
g^{uu} |\partial_u X|^2
- m_X^2 |X|^2
-\frac{\lambda}{4} |X|^4
\right) h_{xy}
$,
and the equation of motion
for $h_{xy}$ can be obtained as
\begin{eqnarray}\label{eq:hxy}
0 = h_{xy}''(u) + \frac{g_1'(u)}{g_1(u)} h_{xy}'(u) + g_2(u) h_{xy}(u)
\end{eqnarray}
with
\begin{eqnarray}
g_1(u) &\equiv& \frac{\sqrt{u}f(u)}{a(u)},
\nonumber \\
g_2(u) &\equiv& g_3(u)
+
\frac{z_0^2 }{4 u f(u)^2} \left( \omega^2 - k^2 f(u) \right)
, \nonumber \\
g_3(u) &\equiv&
-
\frac{1}{2 u a(u) f(u)}
\bigg[
16 u f(u) a''(u)
+8 a'(u) \big( 2 u f'(u)+f(u) \big)
+ a(u) \big(s 2 u f''(u)+f'(u) \big)
%\right.
\nonumber\\
&&
\left.
-6 z_0^2 a(u)^3
+
\frac{ \kappa^2 z_0^2 a^2(u) }{4 u f(u)}T_{xy}(u)
\right].
\end{eqnarray}
Here, 
$g_3(u)$ represents the constant part
toward $\omega$ and $k$
in the potential term $g_2(u)$.
It is seen that
$g_3(u)$ is the obstacle
to use the double integral
dictated above Eq.(\ref{eq:U1F}),
which is similar to the case of Eq.(\ref{eq:W1}).
%---
Thus, 
we may adopt the same analysis as the one for Eq.(\ref{eq:W1}),
namely exploiting the factorization,
\begin{eqnarray}\label{facto_h}
h_{xy}(u) = \chi(u) \phi(u),
\end{eqnarray}
and vanishing the extra term $g_3(u)$.

Under the factorization (\ref{facto_h}),
we can rewrite Eq.(\ref{eq:hxy}) as
\begin{eqnarray}\label{eq:hxy2}
\big(g_1(u) \chi^2(u) \phi'(u)\big)'
+
g_1(u) \chi^2(u)
\left\{
\Psi(u)
+
\frac{z_0^2 }{4 u f(u)^2}\left( \omega^2 - k^2 f(u) \right)
\right\} \phi(u)
=0
\end{eqnarray}
with
\begin{eqnarray}\label{Psi}
\Psi(u)
\equiv
\chi''(u)
+
\frac{g_1'(u)}{g_1(u)} \chi'(u)
+
g_3(u) \chi(u).
\end{eqnarray}

Then, 
$\chi(u)$ is obtained by requiring $\Psi(u)$ to satisfy $\Psi(u)=0$.
However, it can be shown that 
it is difficult
to solve the equation $\Psi(u)=0$ fully.
Similarly to the previous section, 
as the needed solution
eventually is the one in the vicinity of the horizon and the boundary,
we may consider the solution of $\chi(u)$
only in the vicinities at $u=0$ and $u=1$.
%---
Nevertheless, from the reason mentioned below,
one cannot obtain analytically the solution in the expansion around $u=1$.
\newline

Then, 
we will try 
to obtain the solution of $\chi(u)$ 
in the vicinity of $u=0$. 
%---
To this purpose, 
expanding the constant part $\Psi(u)$ 
%in the potential term 
%given of Eq.(\ref{Psi}) 
around $u=0$, 
we will consider the following equation: 
\begin{eqnarray}\label{eq:chi}
\chi'' (u)
+
\left( \frac{1}{u} - 2\mu^2 \right) \chi'(u)
&+&
\left[
-
\frac{1}{u^2}
+
\frac{1}{2u}
\left(
40\mu^2
+
\kappa^2 T'_{xy}(u)\big|_{u=0}
\right)
\right. \nonumber \\ && \quad \left.
+
\frac{1}{4}
\Big\{
-8+\kappa^2
\left(
8\mu^2 T'_{xy}(u)\big|_{u=0}+T''_{xy}(u)\big|_{u=0}
\right)
\Big\}
 \right] \chi(u) = 0.
\end{eqnarray}
To obtain the solution $\chi(u)$ in the vicinity of $u=0$,
we may assume the following form
as the solution of $\chi(u)$,
\begin{eqnarray}\label{eq:chi2}
\chi(u) =
\frac{\chi_{-1}}{u}
+ \chi_{0}
+ \chi_{1} u
+ \chi_{L} \, u  \log(u)
\quad
\textrm{with}
\quad
\chi_{-1} \equiv 1/ z_0^2,
\end{eqnarray}
where $\chi_0$, $\chi_{1}$ and $\chi_L$
are the coefficients to be determined further.
While we have fixed $\chi_{-1}=1/ z_0^2$.
It can be seen that if we assume the form of solution
as in Eq.(\ref{eq:chi2}), then $\chi_{-1}$ can, in general, 
be arbitrary [and each coefficients except for $\chi_{-1}$
can be fixed so as to satisfy Eq.(\ref{eq:chi})]. 
Thus, one should fix $\chi_{-1}$ 
in the first place 
as shown in Ref.\cite{Policastro:2002se}.
By doing so, one can see that
$\chi(u)$ plays the same role
as the background metric $g_{\alpha \alpha}$
at the vicinity of $u=0$, where $g_{\alpha \alpha}(u)$ in the vicinity of $u=0$ behaves as
$\displaystyle g_{\alpha \alpha}(u)=\frac{1}{z_0^2}\frac{1}{u}+\frac{4\mu^2}{z_0^2} +{\cal O}(u)$.
%---
As a consequence, we can put the source of $\phi(u)$ as $h^{(0)}{}_x^y(u)$ with 
$h^{(0)}{}_x^y(u)$ the boundary value.

By substituting $\chi(u)$
with the form in Eq.(\ref{eq:chi2})
and solving the equations
in each order of it,
we can fix the coefficients as
\begin{eqnarray}\label{sol:chi_u0}
\chi_{0} &=&
\frac{\chi_{-1}}{2}
\left(44 \mu ^2+\kappa_2 T_{xy}'(u)\big|_{u=0} \right),\\
%==========
\chi_{1} &=&
\frac{\chi_{-1}}{24 \left(36 \mu ^2+\kappa_2 T_{xy}'(u)\big|_{u=0} \right)}
\bigg[
3 \log (u)
\left(36 \mu ^2+\kappa_2 T_{xy}'(u)\big|_{u=0} \right)
\Big\{1728 \mu ^4+92 \kappa_2 \mu ^2 T_{xy}'(u)\big|_{u=0}
%\right.
\nonumber \\ &&
%\left.
%---
+\kappa_2
\left(
T_{xy}''(u)\big|_{u=0}+\kappa_2 \left(T_{xy}'(u)\big|_{u=0}\right)^2
\right)
\Big\}
-2 \kappa_2
\left\{6 T_{xy}'(u)\big|_{u=0} \left(268 \mu ^4+5 \kappa_2 \mu ^2 T_{xy}'(u)\big|_{u=0}-2\right)
\right.
\nonumber \\ &&
%---
%\left.
+3T_{xy}''(u)\big|_{u=0} \left(54 \mu ^2+\kappa_2 T_{xy}'(u)\big|_{u=0} \right) + 2 T_{xy}^{(3)}(0)\Big\}
-16128 \mu ^6
\bigg], \\
%==========
\chi_{L} &=& -\frac{\chi_{-1} }{8}
\left\{
1728 \mu ^4
+92 \kappa_2 \mu ^2 T_{xy}'(0)
+\kappa_2 \left(
T_{xy}''(u))\big|_{u=0}
+
\kappa_2
\left( T_{xy}'(u)\big|_{u=0}\right)^2\right)
\right\},
\end{eqnarray}
where we have obtained the solution of $\chi(u)$
to the order of $u$, which will become necessary
in the analysis in what follows.

Now, 
we come to discuss the solution $\chi(u)$
in the vicinity of $u=1$. From the expansion of the coefficients of Eq.(\ref{eq:chi})
around $u=1$, one can see that
a general solution in the vicinity of $u=1$ is
given as a linear coupling of the confluent hypergeometric function $U$
and Laguerre polynomials $L_n$.  However, similar to the case of $\rho(u)$
in the previous section, 
it is difficult to fix the integral constants analytically,
so that the constants are common between the ones at $u=1$ and $u=0$
~[Eq.(\ref{eq:chi2}) with the coefficients (\ref{sol:chi_u0})],
due to the lack of the information of the solution of $\chi(u)$ in the interior domain of the bulk.
Thus, 
$\chi(u)$ around $u=1$ will be treated symbolically in the present consideration.
\newline

In the same way as in the previous sections, we can obtain the hydrodynamic solution
for the off-diagonal component of the fluctuation $h_{xy}(u)$ in Eq.(\ref{eq:hxy2})
with $\chi(u)$ determined above. Consequently, it can be written as
\begin{eqnarray}\label{sol:hxy}
h_{xy}(u)
=
\chi(u) \phi(u)
=
C_0^{(\phi)} \, \chi(u) \, (1-u)^{-i \nw/2}
\left(
1
+ \nw F_1^{(\phi)}(u)
+ \nq^2 G_1^{(\phi)}(u)
+ \cdots
\right)
\end{eqnarray}
with
\begin{eqnarray}
C_0^{(\phi)} &=& \frac{h^{(0)}{}^x_y}{1+ G_B^{(\phi)} \nw + F_B^{(\phi)} \nq^2 },\\
%-----
F_1^{(\phi)}(u) &=&
F_B^{(\phi)} - \frac{i}{2}u
+ \frac{i}{2} \left( -\frac{1}{2} + \frac{F_H^{(\phi)}}{z_0^4 \chi_{-1}} \right) u^2,\\
%-----
G_1^{(\phi)}(u) &=&
G_B^{(\phi)}
-
u
+
\left\{
\frac
{
2 \chi_{-1} z_0 \log (u) \left( \chi_0 - \mu^2 \chi_{-1} \right)
+
\chi_{-1} z_0 \left( \chi_0 - \mu^2 \chi_{-1} \right)
}
{2 (\chi_{-1})^2 z_0}
+
\frac{G_B^{(\phi)}}{2 (\chi_{-1})^2 z_0^4}
\right\}
u^2, \nonumber \\
\end{eqnarray}
where $h^{(0)}{}^x_y$ has been mentioned above Eq.(\ref{sol:chi_u0}). $F^{(\phi)}_H$, $F^{(\phi)}_B$ and $G^{(\phi)}_H$, $G^{(\phi)}_B$ represent the integral constants fixed at boundary and horizon, which are similar 
to the integral constants that appeared in Eqs.(\ref{eq:U1F}) and (\ref{eq:U1G}). 
More explicitly, they are fixed as
\begin{eqnarray}
F^{(\phi)}_H &=& \frac{1}{\sqrt{4 \mu ^2+1}}, \\
G^{(\phi)}_H &=& -\frac{1}{\sqrt{4 \mu ^2+1}} G^{(\phi)}_0\big(\mu,\big\{ \Phi^{(n)} \big\}\big).
\end{eqnarray}
Here we have used $\chi_{-1} =1/z_0^2$. It enables us to write down
the solution of $h_{xy}(u)$ in the vicinity of $u=0$ as
\begin{eqnarray}
h_{xy}(u) &=&
\frac{ h^{(0)}{}^x_y \chi_{-1}}{u}
+
h^{(0)}{}^x_y ( \chi_0 - \nq^2 \chi_{-1} )
+
\frac{h^{(0)}{}^x_y}{2 z_0^4 \chi_{-1}^2}
\Big[i F^{(\phi)}_H \omega +q^2
\Big\{
G^{(\phi)}_H + \chi_{-1} z_0^4 \left( \chi_0 - \mu ^2 \chi_{-1} \right)
\Big\} \nonumber \\ &&
+
2 z_0^4 q^2 \chi_{-1}
\left(\chi_0-\mu ^2 \chi_{-1}\right)  \log (u)
\Big]
u
+{\cal O}(u^2) 
\end{eqnarray}
which shows that the expression diverges at $u=0$.
However, such a divergence arises from the contribution of $\chi(u)$
given in Eq.(\ref{sol:chi_u0}), and the solution of $\phi(u)$ itself has been obtained healthily, 
as can be seen from Eq.(\ref{sol:hxy}). Such a situation is the same as the one
discussed in Ref.\cite{Policastro:2002se} and also other papers based on Ref.\cite{Policastro:2002se}.
\newline

Now that 
we have obtained 
the solution, 
let us evaluate the viscosity 
through Kubo formula. 
%---
The boundary action $S_0$ is given by 
\begin{eqnarray}
S_0=S_{\textrm{on-shell}}+S_{\textrm{GH}}+S_{\textrm{ct}},
\end{eqnarray}
with
\begin{eqnarray}
S_{\rm GH}
&=&
\frac{1}{8\pi G_5} \! \int \! d^4x \sqrt{-g^{(4)}} K, \\
S_{\rm ct}
&=&
-
\frac{1}{8\pi G_5} \! \int \! d^4x \sqrt{-g^{(4)}}
\bigg(\  \frac{3}{l} + \frac{l}{4} R^{(4)} \bigg),
\end{eqnarray}
where  
$S_{\rm GH}$ and $S_{\rm ct} $ denote the Gibbons-Hawking term 
and the counter term respectively, 
and $g^{(4)}_{\mu\nu}$ is the four-dimensional induced metric, 
and $K$ and $R^{(4)}$ mean the extrinsic curvature and the curvature on the boundary
toward our geometry (\ref{metric_dads_01}), respectively.
They are found to be 
\begin{eqnarray}
S_{\textrm{on-shell}}
&=&
\frac{1}{2 \kappa^2 z_0}
\int
\frac{dw\, dq}{(2\pi)^2}
e^{-\Phi(u)}
\frac{\sqrt{u}}{a(u)^2}
\Big\{
4 a'(u) f(u) h_{xy}(u)^2 -3 a(u) f(u) h_{xy}(u) h_{xy}{}'(u)
\Big\}
\Big|_{u=0}
,\\
S_{\textrm{GH}}
&=&
\frac{1}{2 \kappa^2 z_0}
\lim_{u\to 0}
\int
\frac{dw\, dq}{(2\pi)^2}
e^{-\Phi(u)}
\frac{\sqrt{u}}{a(u)} \Big\{h_{xy}(u)^2 f'(u)+4 f(u) h_{xy}(u) h_{xy}{}'(u)\Big\},
\\
S_{\textrm{ct}} &=&
\frac{1}{8 \kappa ^2}
\lim_{u\to 0}
\int
\frac{dw\, dq}{(2\pi)^2}
e^{-\Phi(u)}
\frac{12 a(u)^2 f(u) + k^2 f(u) - \omega^2}{ a(u)^2 \sqrt{f(u)}}h_{xy}^2(u).
\end{eqnarray}

The retarded Green function is obtained by using the GKP-W relation 
\begin{eqnarray}
G^{\rm (R)}_{xy \ xy}(\omega ,k)
=
-\frac{N^2T^2}{16}
\Big\{
G^{(\phi)}_H k^2
+2\pi T
\left( i F^{(\phi)}_H \omega - \chi_{-1}^2 \right)
\Big\}.
\end{eqnarray}
which allows us to evaluate the shear viscosity via the Kubo formula
\begin{eqnarray}
\eta
= -\lim_{\omega \to 0} \frac{{\rm Im}\left(G^{\rm (R)}_{xy \ xy}(\omega,k=0)\right)}{\omega}
= \frac{\pi}{8} F_H N^2T^3.
\end{eqnarray}
With this result, it is then not difficult to obtain the ratio 
\begin{eqnarray}\label{etaos}
\frac{\eta}{s}= \frac{1}{4\pi \sqrt{1+4\mu^2}(1+16\pi^2 \mu^2 T^2)^{3/2}},
\end{eqnarray}
where the entropy density is given in Eq.(\ref{TH_and_s}).

\begin{figure}
\begin{center}
\includegraphics[width=70mm,clip]{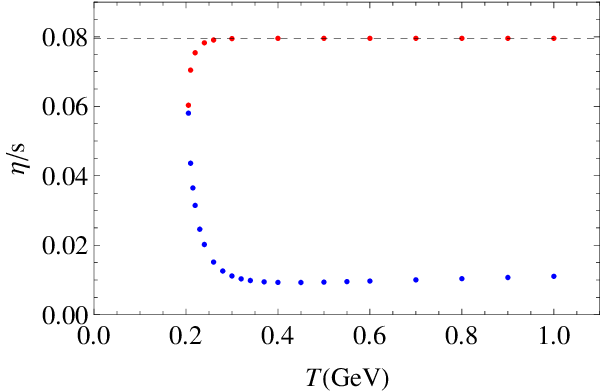}
\includegraphics[width=70mm,clip]{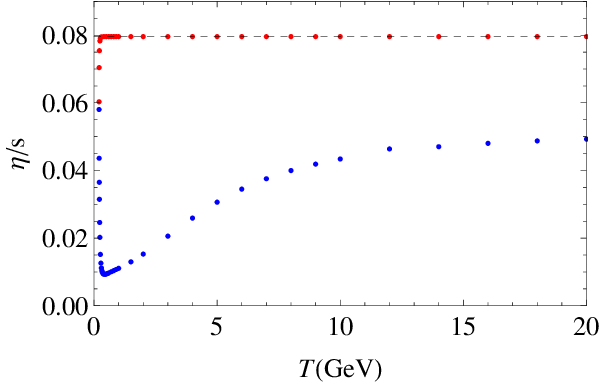}
\end{center}
\caption{
We plot $\eta/s$
given in Eq.(\ref{etaos})
using $\mu$ given in Fig.\ref{Fig_mu}.
%---
%Above two and below two figures represent
%the results
%in the model IIa and IIb respectively,
%and 
The left and right plots are different
just in the scale of the $x$ axis. 
%---
The red and blue points represent the results
obtained
by using the $\mu$
in Fig.\ref{Fig_mu}.
These plots and the plots in Fig.\ref{Fig_mu} correspond each other.
%---
The dashed horizontal line represents the KSS bound, $1/(4\pi)$~\cite{KSSBound}.
%---
In these plots,
the result represented
by the red points 
%in the below two~(the model \textcolor{red}{IIb})
corresponds to
the results 
with regard to the $\mu$
we have chosen
as the physically acceptable one.
}
\label{Fig_etaos}
\end{figure}

The ratio $\eta/s$ 
as the function of temperature is shown 
in Fig.\ref{Fig_etaos}, 
where the left and the right plots are different 
in the scale of the $x$ axis.
%---
The red points and  the blue points represent 
the results  
obtained by using the $\mu$  
in Fig.\ref{Fig_mu}, 
and each plot in Figs.\ref{Fig_etaos} and \ref{Fig_mu} 
corresponds each other. 
%---
The dashed horizontal line represents 
the KSS bound, $1/(4\pi)$~\cite{KSSBound}.
%---
It can be seen that 
there are two branches 
with regard to the solutions of $\mu$ 
in Fig.\ref{Fig_mu}. 
%---
Among the red points and the blue points, 
the red points are the ones that 
we have chosen as the physically acceptable branch
in the last part of Sec.\ref{Chap:AdSQCD model}. 
% , 
%which is represented 
%by the red point. 

%As it has been discussed 
%in the Chap.\ref{Chap:AdSQCD model}, 
% is given 
%by 

%
%The above two and below two plots in Fig.\ref{Fig_etaos} represent
%the results corresponding to the model IIa and IIb, respectively,
%and the left plots and right plots are different just in the scale of $x$ axis.

%---

%============================================
\section{Conclusions and Remarks}
\label{Chap:Sum}
%============================================

In this paper, 
we have extended an IR-deformed AdS/QCD model presented 
in Ref.\cite{Sui:2009xe} 
to a finite temperature system 
by a way 
different from the one proposed in Ref.\cite{Cui:2011ag}, 
worked out the hydrodynamics, 
and computed the transport coefficients. 
%--- 
In the actual analysis,   
we have found several branches, 
and we have chosen the branch 
in which our model can be consistent 
as a holographic model.

The things 
characterizing our model 
are the parameters $\mu_g$, $g_{SU(2)}$, $\lambda$, 
the scalar field $X$ and the dilaton $\Phi$. 
%---
Among these, 
the ones entering our analysis in this paper have been $\mu_g$ 
through $\mu \equiv \mu_g /(2\pi T)$ and dilaton. 
We have treated the effects of these in a symbolic way.  
%---
The remaining things will enter 
if one considers 
the background dual to the system at the finite chemical potential. 
%---
However, 
the analysis with the finite chemical potential
will be too complected to perform the hydrodynamic analysis. 
\newline
%---
%Actually, 
%why we have not performed the scalar mode 
%in the hydrodynamics of the $SU(2)$ flavor current 
%has also been that 
%the term with $g_{SU(2)}$ has made the equation of motion 
%into mostly unsolved form. 
%---
%This matter would be a great obstacle, 
%when we try to the study 
%in our model 
%with finite chemical potential.  

%It is seen that the model involves the quantities $\mu_g$, $g_{SU(2)}$, $\lambda$, the scalar field $X$ and dilaton $\Phi$. While only two of them $\mu_g$ (thorough $\mu \equiv \mu_g /(2\pi T)$ ) and dilaton enter into our analysis in this paper, their effects have actually been treated in a symbolic solution. Note that we have not considered
%the effect from the non-zero background dual to the system at finite chemical potential 
%as it will make the hydrodynamic analysis more complicated.
%Also we have not carried out the analysis on the scalar mode
%in the hydrodynamics of the $SU(2)$ flavor current as
%the term with $g_{SU(2)}$ has made the equation of motion
%into mostly unsolved form. This matter would become a great obstacle
%when one tries to the study with finite chemical potential.

One of the further directions of this study will be 
the holographic description of the Bjorken flow~\cite{Bjorken:1982qr}. 
%-----
The Bjorken Flow is an effective model 
having been invented 
to describe evolution of quark-gluon plasma 
produced in high energy collision experiment. 
%-----
This has the property of the ideal fluid,  
and diffuses with the form of distribution 
in the boost-invariant way 
from a point in Minkowski space-time 
in which the collision happens. 
%-----
It is considered that 
the quark-gluon plasma 
being produced and diffusing in the RHIC or LHC, etc 
can be described by using the Bjorken Flow.  
%-----
However, 
it is known that 
understanding of the Bjorken Flow  
in the framework of the field theory is hard 
due to its strong coupling and the real-time evolution.  
%-----
In such a situation, 
the gauge/gravity correspondence would be useful.  
For the relevant studies, 
we would like to refer the reader 
to the reference in the review papers~\cite{ReviewBjorkenFlow}. 
%-----
In the analyses of these, 
we can see that 
hydrodynamics is a necessity.  
Therefore, 
developing the hydrodynamic analysis performed in this paper more,  
we are going to try to the Bjorken flow in our model.

%It would be interesting to study the holographic description of Bjorken flow~\cite{Bjorken:1982qr}, which is an effective model 
%to \textcolor{red}{describe} the evolution of quark-gluon plasma produced in high energy collision experiment.
%It is considered that the quark-gluon plasma produced and diffusing in RHIC or LHC may well be described by using Bjorken Flow \cite{ReviewBjorkenFlow}, where a %hydrodynamic analysis has been shown to be necessity. Therefore, the hydrodynamic analysis formulated in this paper will enable us to study the Bjorken flow in the %predictive soft-wall AdS/QCD.

\vspace*{5mm} 

\noindent{\large{\bf Acknowledgments}}
The author S.T. would like to thank 
Y. Matsuo,  S. Okazawa, T. Azuma, Y. Imamura and J. P.~Shock 
for useful discussions and the referee for his comment. 
This work is supported in part the National
Nature Science Foundation of China (NSFC) under Grants No. 10975170,
No. 10821504 and No. 10905084 and the Project of Knowledge Innovation
Program (PKIP) of the Chinese Academy of Science.

\end{document}